\documentclass[twocolumn,twocolappendix]{aastex63}

\usepackage{bm}
\usepackage[T1]{fontenc}

\newcommand{\fice}{f_\mathrm{ice}}
\newcommand{\tauice}{\tau_\mathrm{ice}}
\newcommand{\delp}{\Delta{P}}

\newcommand{\amax}{a_\mathrm{max}}
\newcommand{\amin}{a_\mathrm{min}}
\newcommand{\amean}{a_\mathrm{c}}
\newcommand{\water}{\mathrm{H}_2\mathrm{O}}

\newcommand{\pcont}{P^\mathrm{cont}_{\water}}

\turnoffediting

\shorttitle{Large water-ice grains in the outer disk of HD 142527}
\shortauthors{Tazaki et al.}

\begin{document}

\title{The water-ice feature in near-infrared disk-scattered light around HD 142527: Micron-sized icy grains lifted up to the disk surface?}

\correspondingauthor{Ryo Tazaki}
\email{r.tazaki@uva.nl}

\author[0000-0003-1451-6836]{Ryo Tazaki}
\affil{Anton Pannekoek Institute for Astronomy, University of Amsterdam, Science Park 904, 1098XH Amsterdam, The Netherlands}
\affil{Astronomical Institute, Graduate School of Science,
Tohoku University, 6-3 Aramaki, Aoba-ku, Sendai 980-8578, Japan}

\author[0000-0001-9843-2909]{Koji Murakawa}
\affil{Institute of Education, Osaka Sangyo University, 3-1-1 Nakagaito, Daito, Osaka 574-8530, Japan}

\author{Takayuki Muto}
\affil{Division of Liberal Arts, Kogakuin University, 1-24-2 Nishi-Shinjuku, Shinjuku-ku, Tokyo 163-8677, Japan}

\author[0000-0002-6172-9124]{Mitsuhiko Honda}
\affil{Faculty of Biosphere-Geosphere Science, Okayama University of Science, 1-1 Ridai-chou, Okayama 700-0005, Japan}

\author[0000-0002-7779-8677]{Akio K. Inoue}
\affil{Department of Physics, School of Advanced Science and Engineering, Faculty of Science and Engineering, Waseda University, 3-4-1, Okubo, Shinjuku, Tokyo 169-8555, Japan}
\affil{Waseda Research Institute for Science and Engineering, Faculty of Science and Engineering, Waseda University, 3-4-1, Okubo, Shinjuku, Tokyo 169-8555, Japan}




\begin{abstract}
We study the $3~\mu$m scattering feature of water ice detected in the outer disk of HD 142527 by performing radiative transfer simulations. We show that an ice mass abundance at the outer disk surface of HD 142527 is much lower than estimated in a previous study. It is even lower than inferred from far-infrared ice observations, implying ice disruption at the disk surface. Next, we demonstrate that a polarization fraction of disk-scattered light varies across the ice-band wavelengths depending on ice grain properties; hence, polarimetric spectra would be another tool for characterizing water-ice properties. Finally, we argue that the observed reddish disk-scattered light is due to grains with a few microns in size. To explain the presence of such grains at the disk surface, we need a mechanism that can efficiently oppose dust settling. If we assume turbulent mixing, our estimate requires $\alpha\gtrsim2\times10^{-3}$, where $\alpha$ is a non-dimensional parameter describing the vertical diffusion coefficient of grains. Future observations probing gas kinematics would be helpful to elucidate vertical grain dynamics in the outer disk of HD 142527.
\end{abstract}



%
\section{Introduction}\label{sec:intro}
Water ice is the most abundant volatile in the outer regions of protoplanetary disks \citep{Pollack94} and plays a vital role in many aspects of planet formation. First of all, water ice provides a site for efficient chemical reactions, facilitating the formation of complex organic molecules \citep{Herbst09}. Second, ice particles may be sticky enough to promote grain growth \citep{Dominik97, Wada09, Wada13}, although recent studies point toward less sticky properties than previously thought \citep{Gundlach18, Musiolik19, Kimura20}. Third, icy pebbles at the outer disk region may drift radially inward and supply water to terrestrial planets \citep{Sato16, Ida19}. Thus, water ice is essential in various stages of planet formation, and observational studies of ice in the disks will lead to a better understanding of planet formation.

The presence of water ice in protoplanetary disks has been confirmed by observing solid-state features in infrared (IR) wavelengths. At near-IR wavelengths, the $3~\mu$m feature, attributed to the O-H vibration of ice, has been detected as either extinction \citep{Pont05,Terada07,Terada12a,Terada12b,Terada17} or scattering \citep{Honda09,Honda16}. At far-IR wavelengths, the emission features associated with ice lattice modes have been detected with ISO/LWS \citep{Malfait98,Malfait99,Chiang01} and Herschel/PACS \citep{McClure12,McClure15,Min16}. 

Among these ice features, we focus on the 3 $\mu$m scattering feature as it offers a unique opportunity to directly probe the spatial distribution of water ice. The 3 $\mu$m extinction feature has been detected only for nearly edge-on disks \citep{Terada17}, making retrieval of radial ice distribution complicated. In contrast, the far-IR features are observable for nearly face-on disks; however, the limited angular resolution inhibits spatially resolved observations. As the $3~\mu$m scattering feature is at near-IR wavelengths, we can utilize either a ground-based or space-based telescope to achieve a spatially resolved imaging of ice in disks.

The scattering feature is due to the wavelength dependence of the albedo of ice particles \citep{Inoue08}. At a wavelength of approximately 3 $\mu$m, strong absorption due to O-H vibration lowers the albedo of ice particles, resulting in a dimmer disk in scattered light. Thus, the scattering feature appears as an apparent absorption feature in the scattered light spectrum, as detected in HD 142527 \citep{Honda09} and HD 100546 \citep{Honda16}.

HD 142527 is a binary system consisting of two pre-main-sequence stars \citep[e.g.,][]{Lacour16}. The total mass of the binary is $\sim2.2~M_\sun$ ($1.8~M_\sun$ and $0.4~M_\sun$ for primary and secondary stars) with a semimajor axis of $\sim30-50$ au \citep[][references therein]{Price18}.
The strong near-IR excess indicates the presence of the inner dust component likely surrounding the primary star \citep{Avenhaus17}. 
The outer disk, or the circumbinary disk, exhibits intricate structures in near-IR scattered light: inner cavity, spiral structures, shadow-like dips \citep{Fukagawa06, Casassus12, Casassus13, Rameau12, Rodigas14, Avenhaus14, Avenhaus17, Hunziker21}. \citet{Price18} demonstrated that these structures are likely resulting from the interaction between the central binary and the circumbinary disk. 

For the outer disk of HD 142527, both the $3~\mu$m scattering feature and far-IR emission features have been detected \citep{Malfait99, Honda09, Min16}. 
\citet{Honda09} modeled the observed $3~\mu$m feature using a simple radiative transfer model developed in \citet{Inoue08} and obtained an ice/silicate mass ratio of $\sim2.2$ or even higher. \citet{Min16} conducted spectral modeling of the far-IR emission features and obtained an ice/silicate ratio of $1.6^{+0.9}_{-0.6}$.

The ice/silicate ratio of $\sim2.2$ in \citet{Honda09} is seemingly consistent with the value inferred by the far-IR features within the range of the error. However, this ratio requires a rather ice-rich grain. Even if we assume all oxygen atoms in non-refractory elements form water ice, elemental budget analysis yields an ice/silicate ratio of $\sim2.04$ in bulk composition \citep{Min11}. Moreover, \citet{Oka12} claimed that water ice at the disk surface of HD 142527 might suffer photodesorption by far-ultraviolet photons. 

This `too-ice-rich problem' is presumably due to two simplifications in their modeling: (i) isotropic scattering and (ii) a simplified disk geometry \citep{Inoue08, Honda09}. The reddish disk-scattered light of HD 142527 \citep{Avenhaus14, Hunziker21} points to larger grains so that the isotropic scattering assumption is no longer valid \citep[e.g.,][]{Mulders13}. Once anisotropic scattering comes into play, a disk geometry may also influence the scattering feature as it determines scattering angles. Therefore, the too-ice-rich problem may be attributed to these simplifications.

This paper revisits radiative transfer modeling of the 3 $\mu$m scattering feature of water ice at the outer disk around HD 142527. We aim to determine grain radius and ice abundance in the disk surface by taking anisotropic scattering and a more realistic disk geometry into account. In addition, we discuss the feasibility of observing the ice scattering feature with the Near Infrared Camera (NIRCam) of the James Webb Space Telescope (JWST). We also investigate polarimetric spectra of the scattering feature, which may offer another opportunity of characterizing ice \citep{Pendleton90, Kim19, Tazaki21}.

This paper is organized as follows. In Section \ref{sec:model}, we summarize the dust and disk models used in this study. We first study the total intensity spectra obtained by our radiative transfer simulations in Section \ref{sec:result} and then investigate polarimetric spectra in Section \ref{sec:result2}. Section \ref{sec:disc} discusses the outcomes of our simulations, mainly focusing on ice abundance and vertical grain dynamics. Section \ref{sec:summary} summarizes the principal results.

\section{Models and methods} \label{sec:model}

\subsection{Dust models} \label{sec:dust}

We adopt dust models in our companion paper \citep{Tazaki21}, as summarized below. We assume that each dust grain is composed of silicate and water ice. The fractional mass abundances of silicate and water-ice grains relative to the total gas and solid phases are determined by $\zeta_\mathrm{sil}=\zeta_\mathrm{sil}^\mathrm{P94}$ and $\zeta_\mathrm{ice}=\fice\zeta_\mathrm{ice}^\mathrm{P94}$, where $\zeta_\mathrm{sil}^\mathrm{P94}=2.64\times10^{-3}$ and $\zeta_\mathrm{ice}^\mathrm{P94}=5.55\times10^{-3}$ are the abundances taken from \citet{Pollack94}. 
$\fice$ is a free parameter to adjust the water-ice abundance. The material densities of silicate and ice are 3.5 g cm$^{-3}$ and 0.92 g cm$^{-3}$, respectively. Table \ref{tab:fvol} shows mass (volume) fractions of silicate and ice for various values of $\fice$. We remind the readers that \citet{Honda09} suggested $\fice\gtrsim1$ for HD 142527.

\begin{table}[tbp]
  \caption{Mass Fraction (Volume Fraction) of Ice and Silicate in Each Grain}
  \label{tab:fvol}
  \centering
  \begin{tabular}{lccc}
    \hline
    Model & Water Ice  & Silicate & ice/silicate mass ratio\\
    \hline \hline
    $f_\mathrm{ice}=1$ & 68\% (89\%) & 32\% (11\%) & 2.1 \\
    $f_\mathrm{ice}=0.3$ & 39\% (71\%) & 61\% (29\%) & 0.63\\    
    $f_\mathrm{ice}=0.1$ & 17\% (44\%)& 83\% (56\%) & 0.21 \\
    $f_\mathrm{ice}=0.03$ & 6\% (19\%) & 94\% (81\%) & 0.063 \\
    \hline
  \end{tabular}
\end{table}

We use the Mie theory combined with an effective medium approach to obtain optical properties of an ice/silicate mixture \citep{Bohren83}. The refractive indices of silicate and water ice are taken from \citet{Draine03b} and \citet{Warren08}, respectively. We derive an effective refractive index by using the Bruggeman mixing rule \citep{Bruggeman35}. 

The refractive index of water ice at the 3-$\mu$m band is strongly dependent on temperature and form of ice. Although \citet{Min16} found that far-IR ice emission features observed for HD 142527 are of highly crystalline nature, ice still can be amorphous up to 40 \% in mass. As mentioned, we use the refractive index of crystalline water ice by \citet{Warren08} mainly because its wide wavelength coverage is useful for radiative transfer simulations. Despite its crystalline nature, its ice-band structure is similar to warm amorphous ice (see Appendix \ref{sec:amo}).

We average the optical properties by considering grain-size distribution. We consider two types of grain-size distribution. 
One is a log-normal distribution defined by 
\begin{equation}
n(a)da \propto \exp\left[-\frac{(\ln(a/\amean))^2}{2\sigma^2}\right] d\ln{a},
\end{equation}
where $a$ is the grain radius, $n(a)da$ is the number density of grains with radii between $a$ and $a+da$,  $\amean$ is the mean grain radius, and $\sigma=0.1$.
The other one is a power-law distribution obeying 
\begin{eqnarray}
n(a)da\propto\left\{ \begin{array}{ll}
a^{-q}da & (\amin\le a \le \amax), \\
0 & (\mathrm{otherwise}), \\
\end{array} \right.
\end{eqnarray}
where $q$ is the power-law index, and $\amin$ and $\amax$ are the minimum and maximum grain radii, respectively. We assume $\amin=0.01~\mu$m and $q=3.5$. As long as $\amin\ll\amax$, the result is insensitive to $\amin$. 

We consider three different grain sizes: $0.3~\mu$m, $1~\mu$m, and $3~\mu$m for both $a_c$ and $\amax$. These size distributions are only valid for the disk surface because the disk midplane is dominated by further large grains \citep{Kataoka16, Ohashi18}.

\begin{figure*}[t!]
\begin{center}
\includegraphics[width=\linewidth]{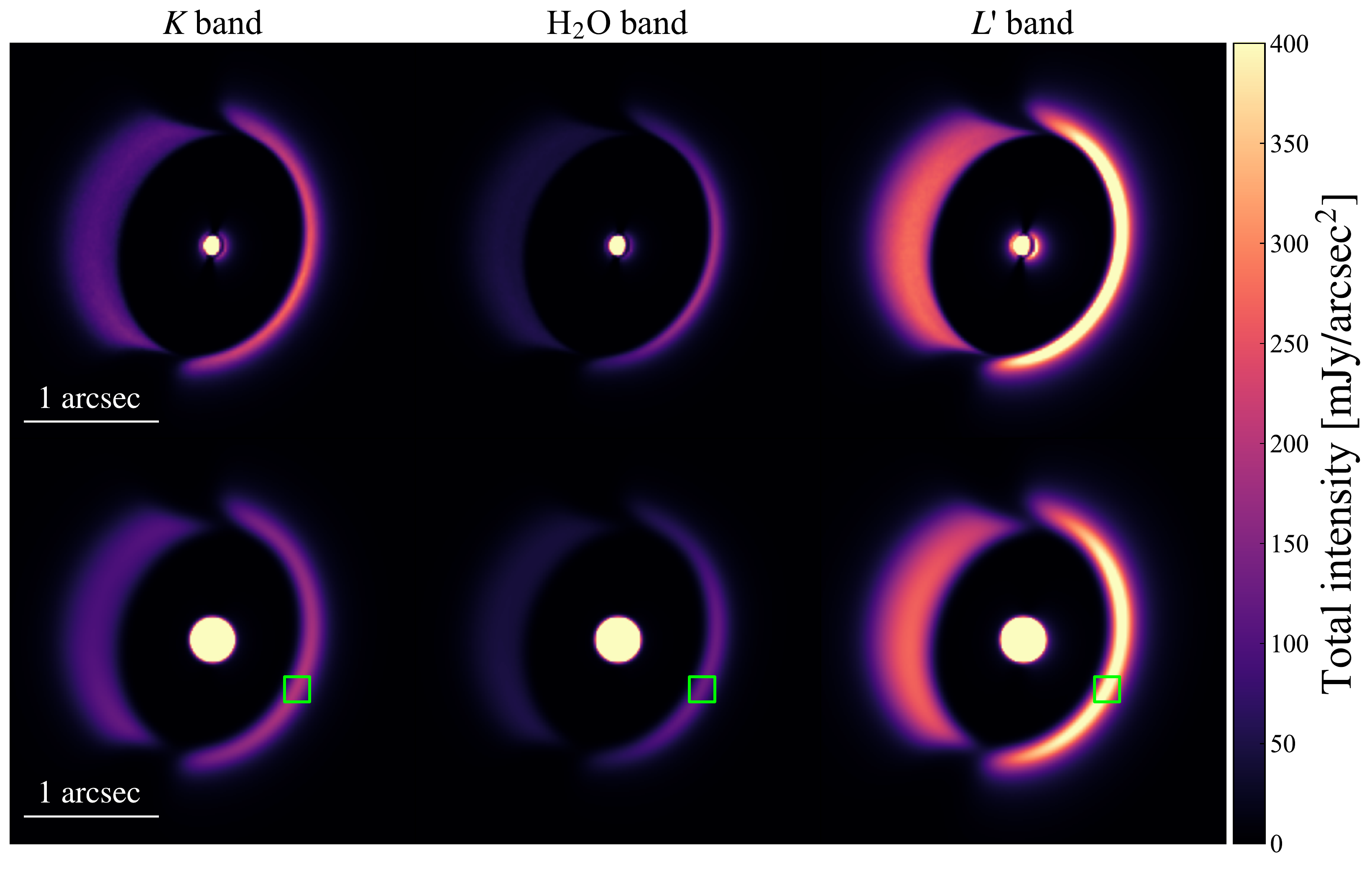}
\caption{Total intensity images of the simulated HD 142527 system at the $K$, $\water$, and $L'$ bands for the log-normal grain-size distribution with $a_c=3~\mu$m and $\fice=0.03$. The two intensity minima around the top and bottom parts of the outer disk are the shadows cast by the misaligned inner disk. Top and bottom panels correspond to images before and after the PSF convolution, respectively. The square symbols in the bottom panels indicate the location at which spectra shown in Figures \ref{fig:spec}, \ref{fig:spec2}, and \ref{fig:cc} were measured. The bottom-right side in each image corresponds to the near side of the outer disk.}
\label{fig:image}
\end{center}
\end{figure*}

\subsection{Star and disk models} \label{sec:rt}

We carry out radiative transfer simulations to investigate the spectral properties of disk-scattered light around HD 142527. Instead of the binary stars, we assume a single star with a radius of $2.5R_\sun$ and an effective temperature of 6250 K \citep{Verhoeff11}. 

A disk model around HD 142527 is taken from \citet{Marino15} with slight modifications. The disk model consists of three components: the inclined inner disk, gap region, and outer disk. The inner disk is inclined by $70^\circ$ with respect to the outer disk. The position angles (PAs) of inner and outer disks are $-8^\circ$ and $-20^\circ$, respectively. Although the inner disk geometry could be more complex than that adopted in \citet{Marino15} as discussed in \citet{Avenhaus17}, we still use this model as it successfully reproduces the total flux spectrum of the HD 142527 system (see Figure \ref{fig:spec}). We modified the outer dust disk mass from $\sim10^{-2}M_\sun$ to $10^{-3}M_\sun$, as inferred from previous studies \citep{Verhoeff11,Min16, Muto15, Boehler17}. Also, we omit the asymmetric azimuthal structure of the outer disk. The inclination angle of the outer disk is 24$^\circ$. Following \citet{Marino15}, the distance of the disk is assumed to be 140 pc, although a more recent study favors a larger distance of 157$\pm1$ pc \citep{Arun19}. 

For the inner disk, we adopt silicate grains obeying the log-normal distribution with $a_c=1~\mu$m so as to reproduce the observed total flux. For the outer disk, we consider grain models presented in Section \ref{sec:dust}, and these grains are assumed to be well mixed throughout the outer disk. To clarify the influence of grain size on ice features, we also assume the same ice abundance across the outer disk, although ice abundance may vary radially and vertically in more realistic disk models \citep{Oka12, Kamp18, Tung20, Ballering21}. This suggests that an ice abundance inferred by our simulations may correspond to the average ice abundance along each line of sight.

We consider band filters from Subaru/IRCS and JWST/NIRCam \footnote{The transmittance data of each band filter is available at 
\url{https://www.naoj.org/Observing/Instruments/IRCS/camera/filters.html} for Subaru/IRCS and \url{https://jwst-docs.stsci.edu/near-infrared-camera/nircam-instrumentation/nircam-filters} for JWST/NIRCam.}. For Subaru/IRCS, we chose three broadband filters: $H$ (1.63 \micron), $K$ (2.20 \micron), and $L'$ (3.77 \micron), and a narrow-band filter $\water$ Ice (3.05 \micron). For JWST/NIRCam, we chose three medium band filters: F250M (2.503 \micron), F300M (2.989 \micron), and F360M (3.624 \micron). The point-spreading function (PSF) of the Subaru and JWST filters is typically around $0\arcsec.1$. We thus convolve an obtained image with the two-dimensional Gaussian function with FWHM of $0\arcsec.1$.

For the given dust and disk models, we perform 3D Monte-Carlo radiative transfer simulations using \texttt{RADMC-3D v2.0} \citep{Dullemond12} with a full scattering polarization treatment. We use the spherical coordinate system. The zenith and azimuthal cells are linearly equispaced with 256 cells. The radial cells are logarithmically equispaced with 256 cells, where half of them are allocated to the inner disk and gap region, and the other half to the outer disk.
We use the photon packets of $10^8$ for both thermal and scattering Monte Carlo simulations. We create disk images from $\lambda=1.4~\mu$m to $4.2~\mu$m with 77 wavelengths and extract scattered light spectrum at each location of the outer disk after the PSF convolution was performed.

\section{Total intensity spectra of disk-scattered light} \label{sec:result}

In this section, we investigate total intensity spectra of disk-scattered light.

\subsection{Grain radius and ice abundance} \label{sec:spec}
Figure \ref{fig:image} shows total intensity images of the simulated HD 142527 system at the $K$, $\water$, $L'$ bands for the log-normal distribution with $a_c=3~\mu$m and $\fice=0.03$. Disk-scattered light at the $\water$ band is found to be dimmer compared to the other two bands, indicating the presence of water ice.
Two intensity minima on the outer disk are the shadows cast by the misaligned inner disk \citep{Marino15}.

\begin{figure}[t]
\begin{center}
\includegraphics[width=\linewidth]{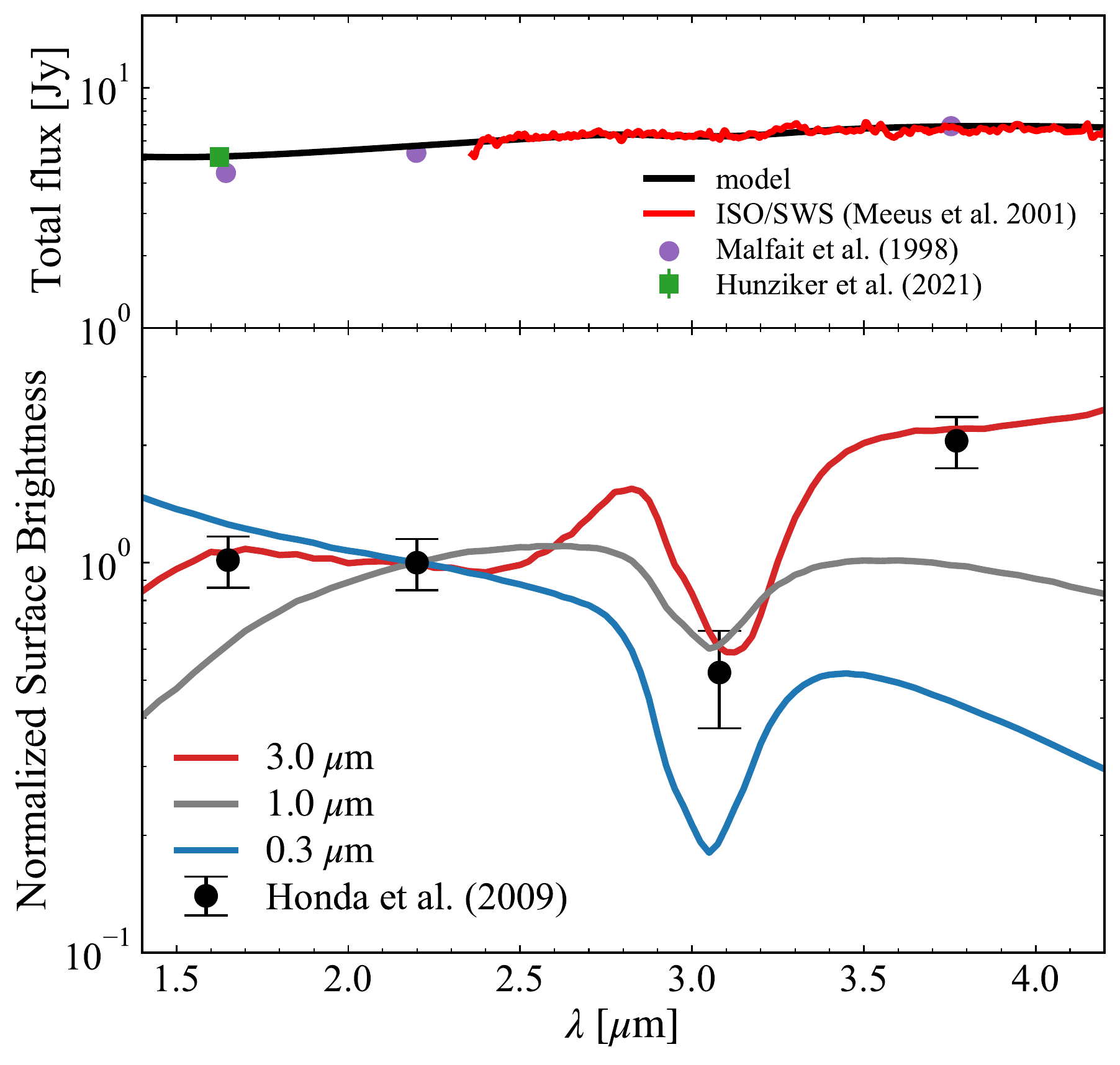}
\caption{(Top) The total flux of HD 142527. The black and red solid lines show the model spectrum and the ISO/SWS spectrum from \citet{Meeus01}, respectively. The green and purple points are photometric flux taken from \citet{Hunziker21} and \citet{Malfait98b}, respectively. (Bottom) The surface brightness of the outer disk at PA=$240^\circ$ normalized at the $K$-band. The blue, gray, and red lines show the model spectra for $a_c=0.3~\mu$m, $1~\mu$m, and $3~\mu$m, respectively. The ice abundance is set as $\fice=0.03$. The black data points are observational data for the region A in \citet{Honda09}.}
\label{fig:spec}
\end{center}
\end{figure}

\begin{figure}[t]
\begin{center}
\includegraphics[width=\linewidth]{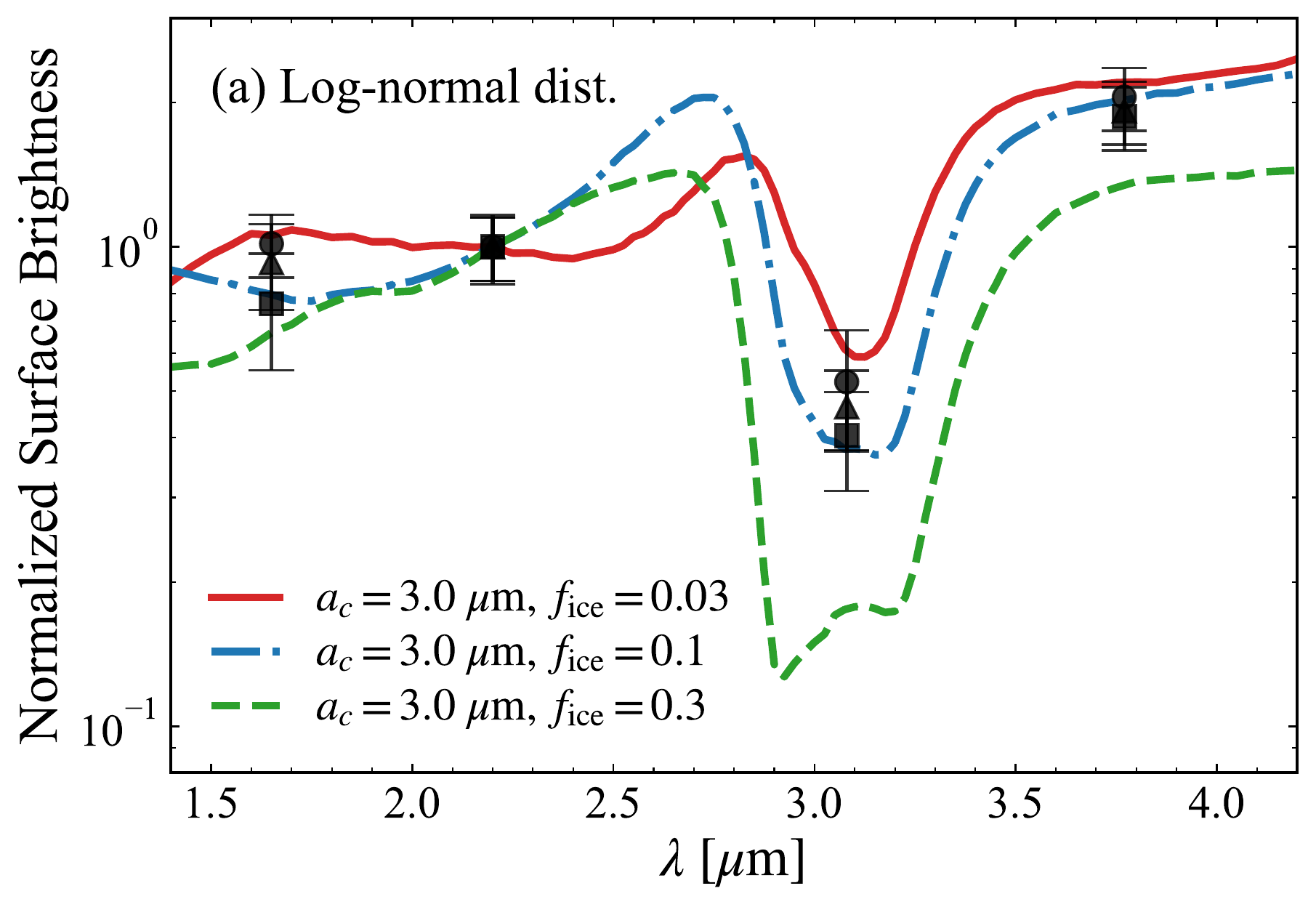}
\includegraphics[width=\linewidth]{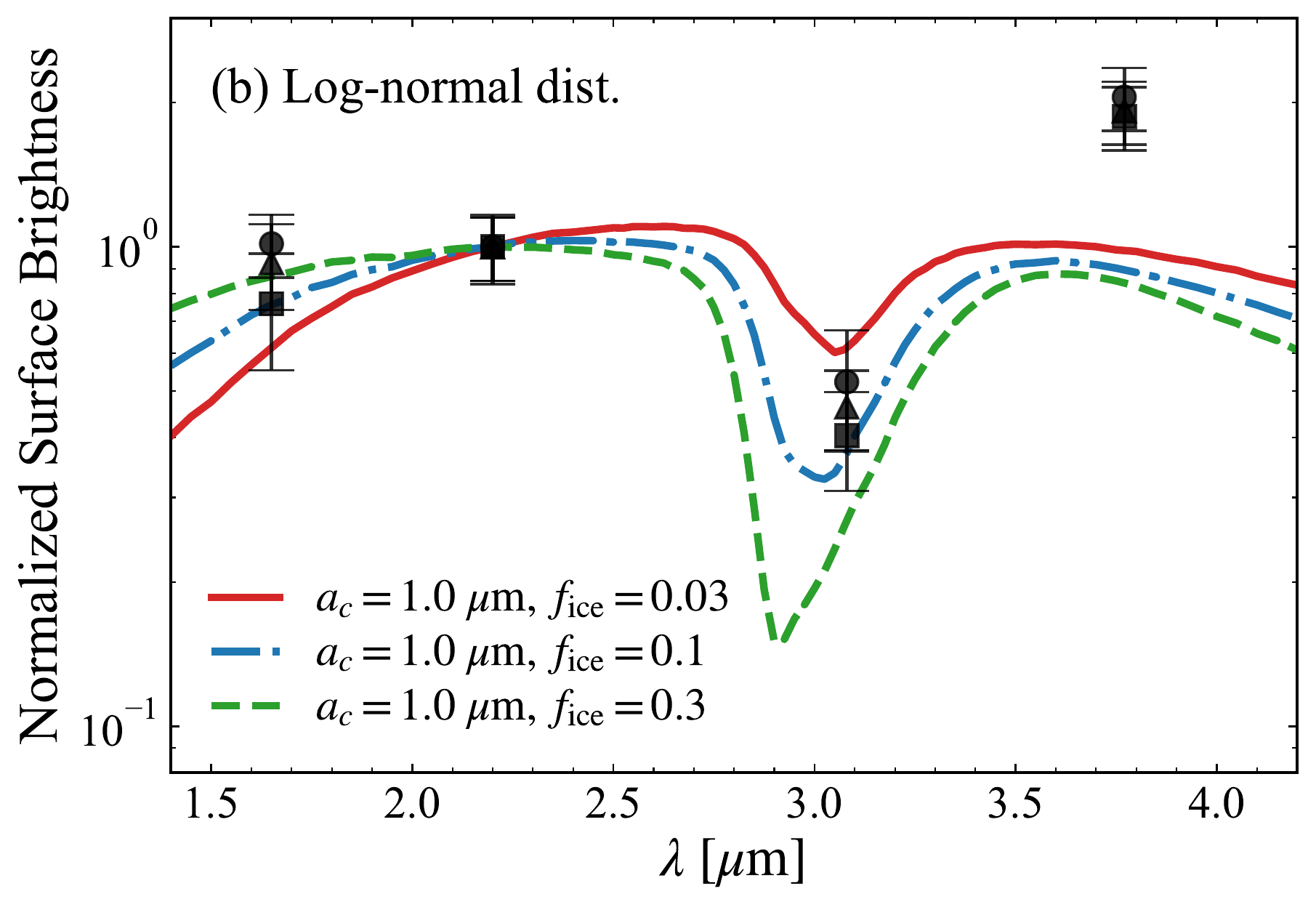}
\includegraphics[width=\linewidth]{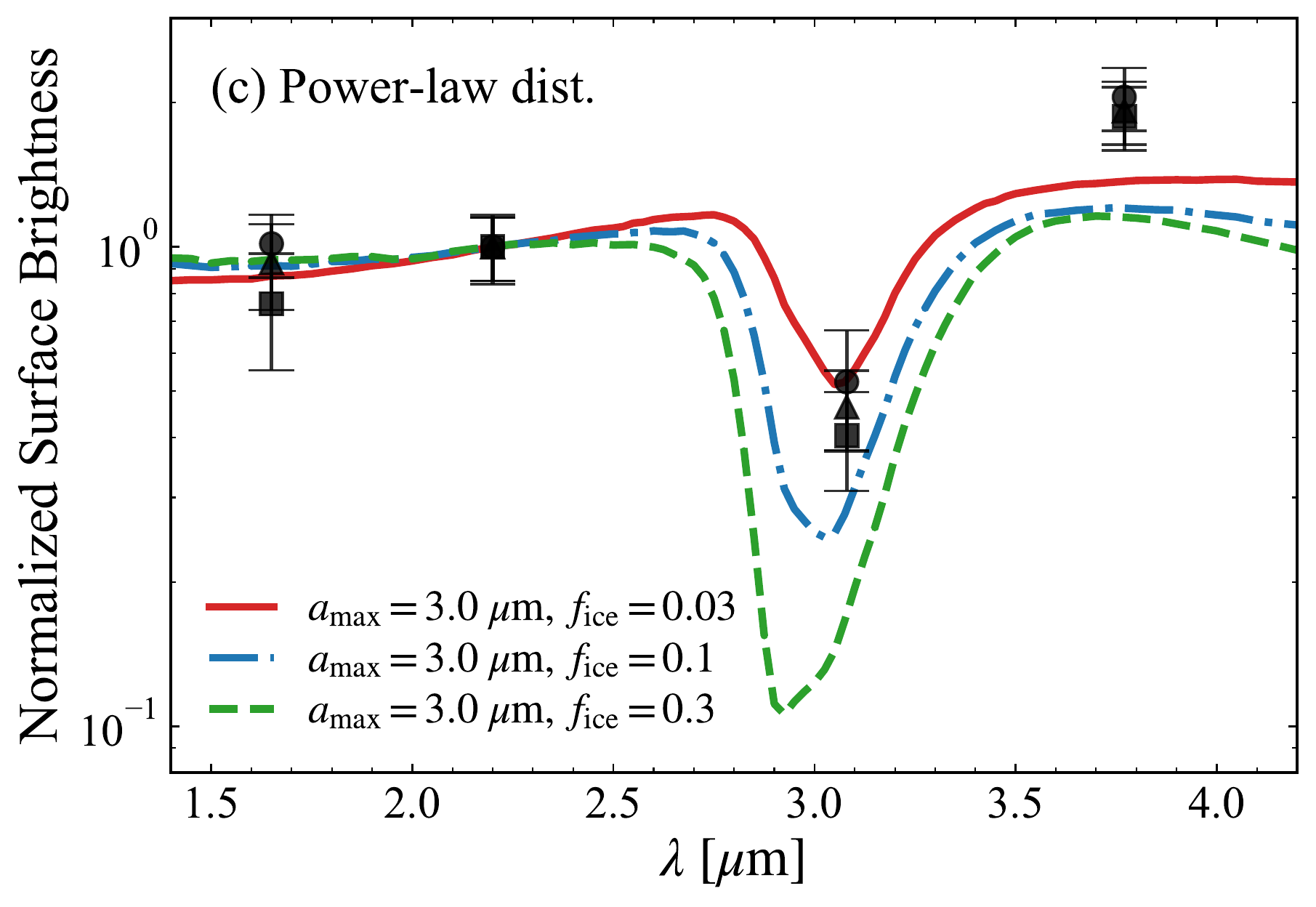}
\caption{(a) Same as Figure \ref{fig:spec}(bottom), but for different ice abundances. The solid, dot-dashed, and dashed lines represent the results for $a_c=3~\mu$m with $\fice=0.03$, $0.1$, and $0.3$, respectively. The circle, triangle, and square symbols represent the observations at the regions A, B, and C in \citet{Honda09}, respectively. (b) Same as (a), but for $a_c=1~\mu$m. (c) Same as (a), but for the power-law distribution.}
\label{fig:spec2}
\end{center}
\end{figure}

Figure \ref{fig:spec}(top) shows the total flux spectrum for the same model shown in Figure \ref{fig:image}. Our model spectrum is quantitatively consistent with both near-IR photometric flux \citep{Malfait98b, Hunziker21} and the ISO/SWS spectrum \citep{Meeus01}. Because the total flux is mostly coming from the inner disk, the spectrum is featureless. 

Figure \ref{fig:spec}(bottom) shows the outer disk-scattered light spectra for small ($a_c=0.3~\mu$m), medium ($1~\mu$m), and large ($3~\mu$m) grains. Following \citet{Honda09}, each spectrum was derived at the location of the peak intensity at the $\water$ band along PA=$240^\circ$, as indicated in Figure \ref{fig:image}. For all model spectra, we can clearly see a dip at $\lambda\sim3~\mu$m, and this is the scattering feature of water ice. With increasing the grain radius, the scattered light color turns from blue to red. The reddish scattered light for the large grains is due to strong forward scattering for shorter wavelengths, which effectively reduces the scattering efficiency of grains \citep{Mulders13, Tazaki19a}. 

We compare these model spectra with the observed surface brightness at the region A in \citet{Honda09}, which corresponds to the peak intensity region along PA=$240^\circ$ at the $\water$ band. Also, Figure \ref{fig:spec2} compares our various model spectra with the observations at the regions A, B, and C in \citet{Honda09}; the regions B and C are located along the same PA as the region A but are measured at larger radial distances. As a result, we find that the observed spectra for all three regions are well reproduced by the log-normal distribution with $a_c=3~\mu$m and $\fice\sim0.03-0.1$. 

Our results point toward larger and ice-poor grains than estimated in \citet{Honda09}, where the authors ascribed the observational results to ice-rich grains ($\fice\gtrsim 1$) of a radius of $\sim 1~\mu$m. Therefore, the too-ice-rich problem has been successfully resolved in our simulations. 

Another intriguing finding here is the presence of large grains at the disk surface. Because such grains are typically anticipated to settle down below the scattering surface, our results hint at strong turbulence or disk winds, which may efficiently inhibit dust settling. We will discuss these points in Sections \ref{sec:alpha} and \ref{sec:lift} in more detail.

\subsection{Ice feature profiles} \label{sec:prof}

We summarize some characteristics of scattering feature profiles obtained in our simulations.

As shown in Figure \ref{fig:spec2}(a), the feature profile for the log-normal distribution with $a_c=3~\mu$m and $\fice=0.03$ peaks at $3.1~\mu$m. With increasing ice abundances, the peak position shifts to shorter wavelengths, i.e., $2.9~\mu$m for $\fice=0.3$. The $2.9~\mu$m peak corresponds to the wavelength at which the real part of the refractive index is minimized (e.g., Figure \ref{fig:opcont}). In addition, the feature becomes double-peaked owing to wavelength dependence of albedo of large icy grains \citep[see also][]{Inoue08, McCabe11}. The double-peaked profile can be an indicator of the presence of large ice-rich grains, as it is not seen for $a_c=1~\mu$m (Figure \ref{fig:spec2}b).

The feature profile also depends on grain-size distribution. Figure \ref{fig:spec2}(c) shows the feature profiles for the power-law distribution with $\amax=3~\mu$m. In this case, the peak position shifts to shorter wavelengths with increasing $\fice$, while the profile remains single-peaked. These properties are similar to those of the log-normal distribution with $a_c=1~\mu$m (Figure \ref{fig:spec2}b). This is because the optical properties of the power-law distribution are mostly dominated by grains of $\sim1~\mu$m in size. 

\subsection{Effect of size distribution on scattered light spectra} \label{sec:sizedist}

\begin{figure*}[t]
\begin{center}
\includegraphics[width=\linewidth]{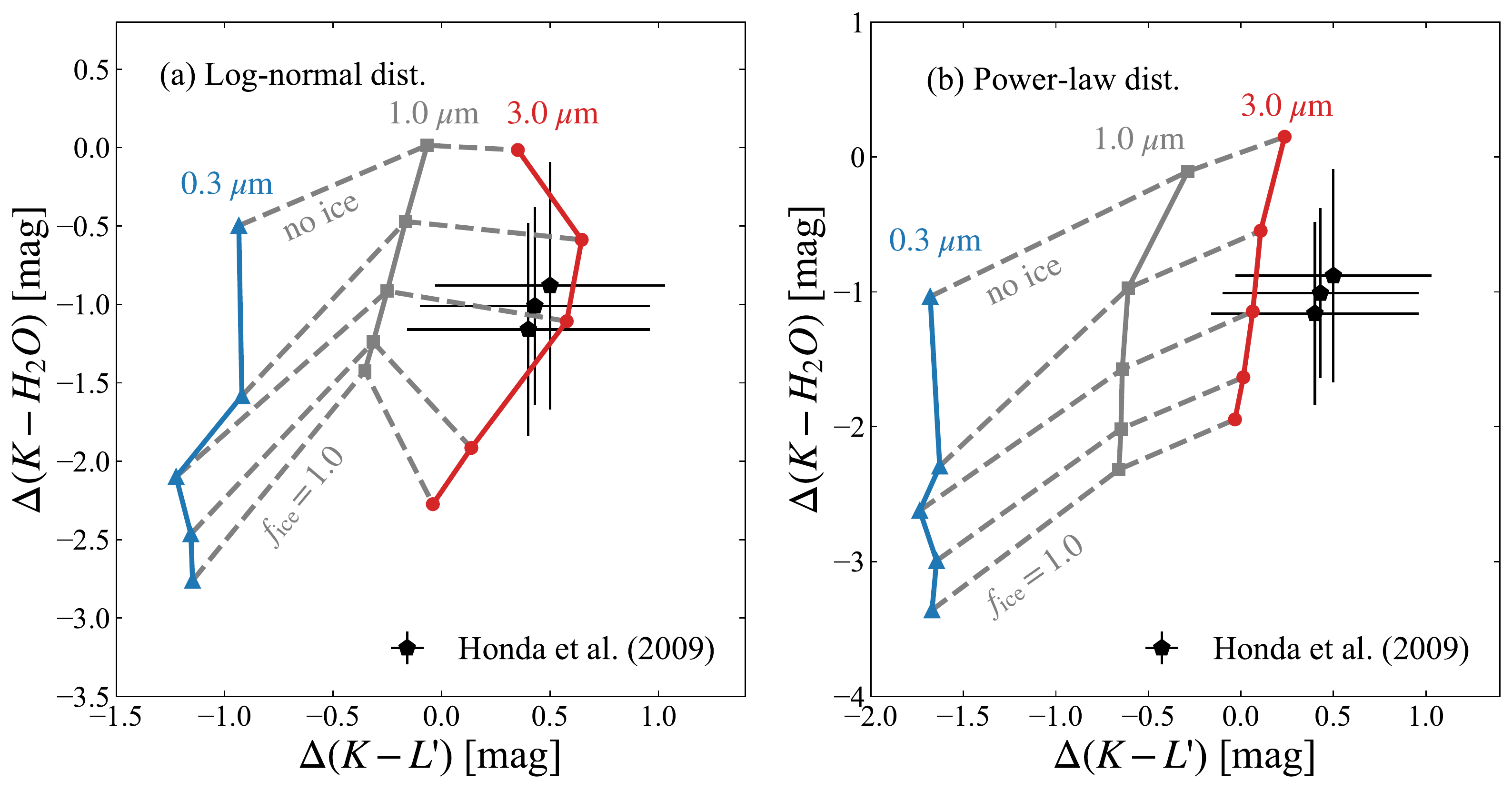}
\caption{The two-color difference diagrams measured at the location indicated in Figure \ref{fig:image}. The left and right panels are the results for the log-normal and power-law distribution, respectively. The triangles, squares, and circles show the results for $a_c$ (left) and $\amax$ (right) of $0.3~\mu$m, $1~\mu$m, and $3~\mu$m, respectively. Each gray dashed line shows the constant $\fice$ line (from bottom to top, $\fice=1.0$, $0.3$, $0.1$, $0.03$, and no ice). The pentagon symbols are the observed values for the regions A, B, and C in \citet{Honda09}.}
\label{fig:cc}
\end{center}
\end{figure*}

To provide a panoptic view on which combination of dust radius and ice abundance can explain the observations, we consider a two-color difference diagram proposed by \citet{Inoue08}. The color difference is defined by
\begin{equation}
\Delta(K-\water)=(K-\water)_\mathrm{sca}-(K-\water)_\mathrm{total},
\end{equation}
where $(K-\water)_\mathrm{sca}$ and $(K-\water)_\mathrm{total}$ are the colors of the outer disk-scattered light and total flux of the system, respectively. Likewise, we define $\Delta(K-L')$ to measure the color difference of the continuum spectrum. Larger grains and shallower ice features tend to yield larger $\Delta(K-L')$ and $\Delta(K-\water)$ values, respectively.

\begin{figure}[t]
\begin{center}
\includegraphics[width=\linewidth]{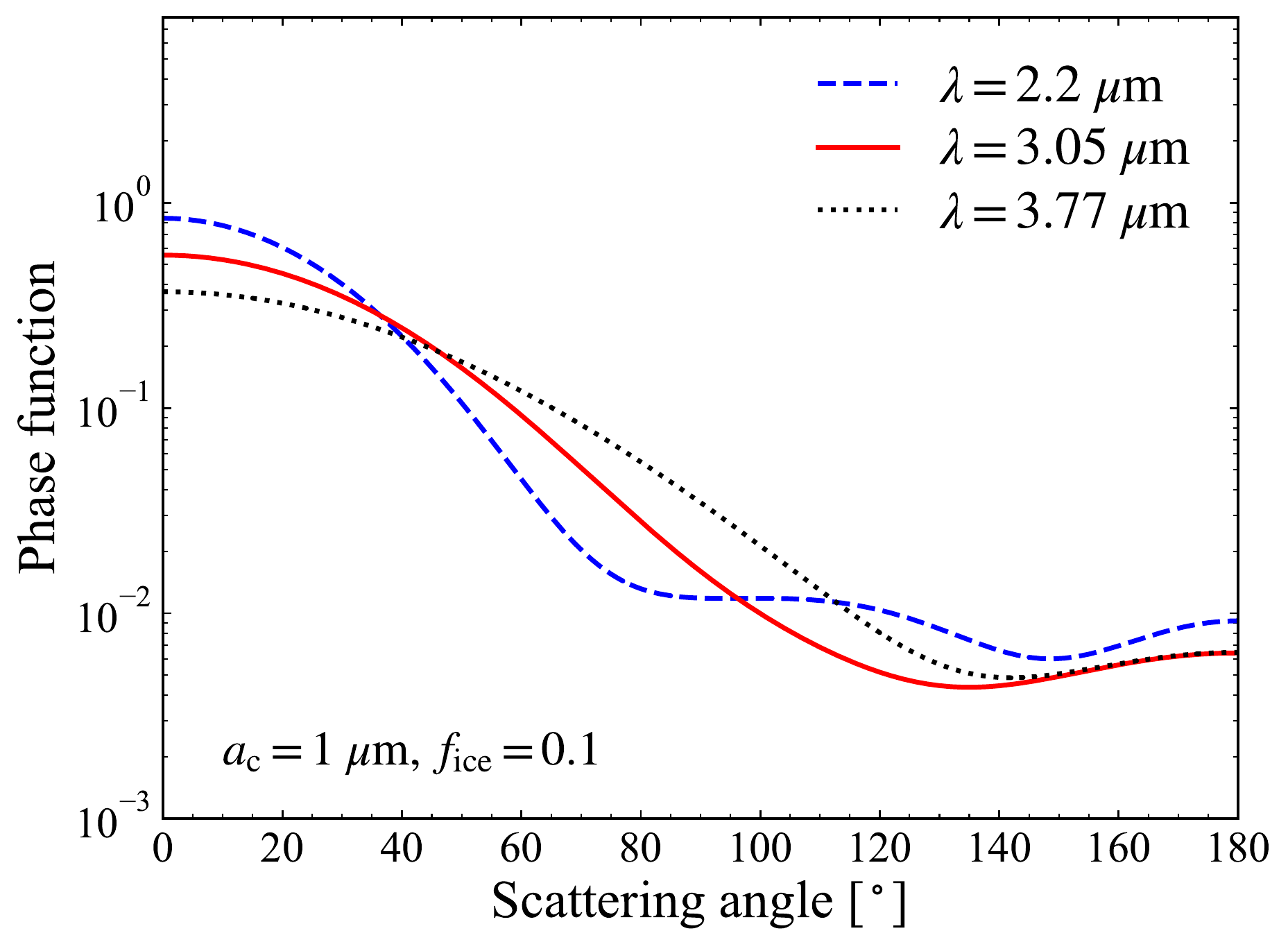}
\includegraphics[width=\linewidth]{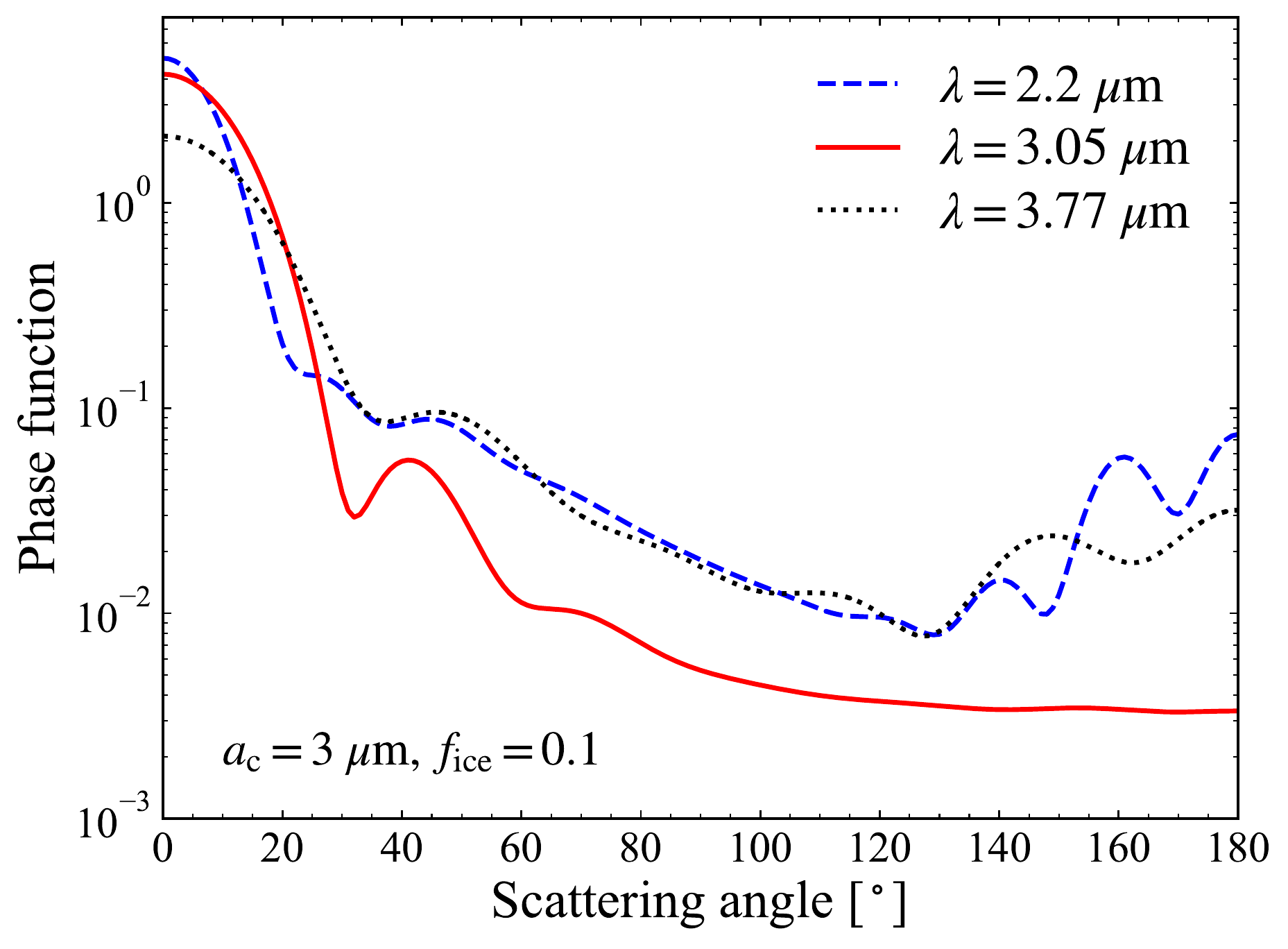}
\caption{Phase functions for the log-normal distribution with $a_c=1~\mu$m (top) and $3~\mu$m (bottom) at three wavelengths across the ice feature. The ice abundance is set as $\fice=0.1$. The red solid line is measured at the $\water$ band wavelength, whereas the other lines are at the $K$ and $L'$ band wavelengths. The observable range of scattering angles is typically $40^\circ-130^\circ$ (see Figure \ref{fig:scatang}).}
\label{fig:phase} 
\end{center}
\end{figure}

\begin{figure}[t]
\begin{center}
\includegraphics[width=\linewidth]{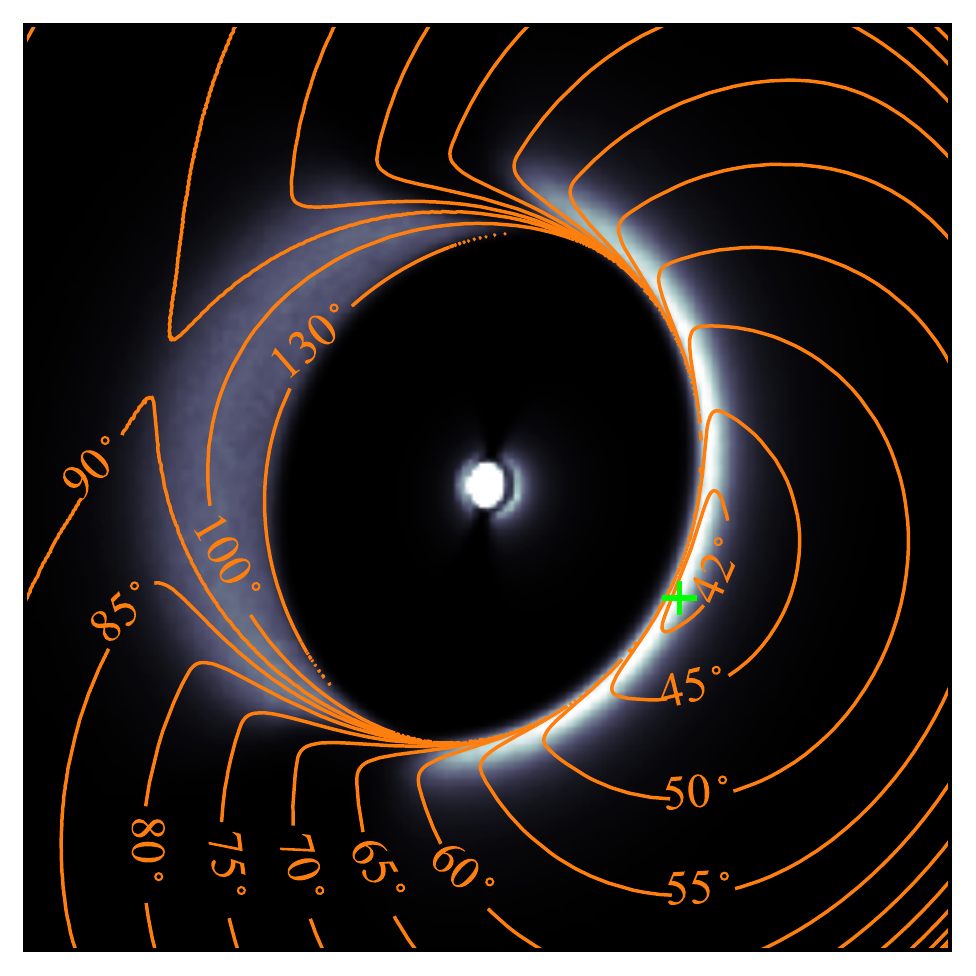}
\caption{A total intensity image at $\lambda=2.2~\mu$m overlaid with the scattering angle contours. The assumed dust model is the log-normal distribution with $a_c=3~\mu$m and $\fice=0.1$. The plus symbol indicates the region at which the spectra shown in Figures \ref{fig:spec}, \ref{fig:spec2}, and \ref{fig:cc} were measured (PA=240$^\circ$). The scattering angles vary from $40^\circ$ (the near side) to $130^\circ$ (the far side). The scattering angle is calculated by ignoring the inner disk for simplicity, and hence, it corresponds to the value when the outer disk is directly illuminated by the central star.}
\label{fig:scatang}
\end{center}
\end{figure}

Figure \ref{fig:cc} shows the color-color diagram for two types of grain-size distribution. Both types of distribution require a similar ice abundance $\fice\sim0.03-0.1$ to explain the observations. The observed reddish scattered light, $\Delta(K-L')>0$, is better captured by the log-normal distribution (see also Figure \ref{fig:spec2}). This is because, for the case of power-law distribution, the scattering properties for $\amax=3~\mu$m are mostly dominated by $\sim1~\mu$m grains in the distribution, and then, the scattered light spectra become almost gray. The gray color remains almost the same for further larger $\amax$. 

Although the power-law distribution models are consistent with the observations within the $1\sigma$ error bars, our results slightly favor the log-normal distribution models. If this is true, grain-size distribution at the disk surface is suggested to be narrow. In other words, a top-heavy (smaller $q$) or a narrow (larger $\amin$) power-law distribution than our model ($q=3.5$ and $\amin=0.01~\mu$m) would provide a closer match to the observations. In this sense, at the disk surface, sub-micron-sized grains might be underabundant relative to micron-sized grains.

In Figure \ref{fig:cc}(a), the log-normal distribution models exhibit the characteristic behavior; $\Delta(K-\water)$ increases from $a_c=0.3~\mu$m to $1~\mu$m, but decreases from $1~\mu$m to $3~\mu$m. This means that the ice feature for $a_c=3~\mu$m can be deeper than that for $1~\mu$m. This behavior is absent in the power-law distribution models as well as isotropic scattering models in previous studies \citep{Inoue08, Honda09}.

We ascribe this behavior to wavelength dependence of phase function. Figure \ref{fig:phase} shows phase functions for $a_c=1~\mu$m and $3~\mu$m at wavelengths across the ice feature. For comparison, Figure \ref{fig:scatang} shows scattering angles for the outer disk, which typically range from $\sim40^\circ$ (near side) to $\sim130^\circ$ (far side). The spectra were measured at a location with a scattering angle of $\sim41^\circ$. 

For $a_c=3~\mu$m, the phase function at the $\water$ band wavelength is significantly different from those at the other wavelengths, whereas such difference is not seen in the case of $1~\mu$m. This is because large icy grains are highly absorbing at the $\water$ band so that they can readily attenuate multiple internal scattering, rendering large-angle scattering being suppressed \citep[see e.g., Figure 12 in][]{Tazaki21}. Due to this difference in phase functions, the ice feature for $a_c=3~\mu$m can be deeper than that for $1~\mu$m grains. The difference also suggests that the feature can be deeper for a lager scattering angle, as shown in Section \ref{sec:tauice}. Forward scattering tends to be enhanced as well at $\lambda\sim2.9~\mu$m, where the real part of the refractive index of the ice band diminishes.

Therefore, the wavelength dependence of phase function has a role in making the feature deeper, which decreases $\Delta(K-\water)$ values for large grains. As the isotropic scattering model ignores this effect, a resultant ice abundance would be overestimated. This is the reason why our inferred abundance comes at lower than that inferred by \citet{Honda09}. This effect is also absent for the power-law distribution with $\amax=3~\mu$m as $1~\mu$m grains tend to govern its phase functions.

\subsection{Dependence of ice feature depth on disk regions} \label{sec:tauice}

\begin{figure*}[t]
\begin{center}
\includegraphics[width=0.49\linewidth]{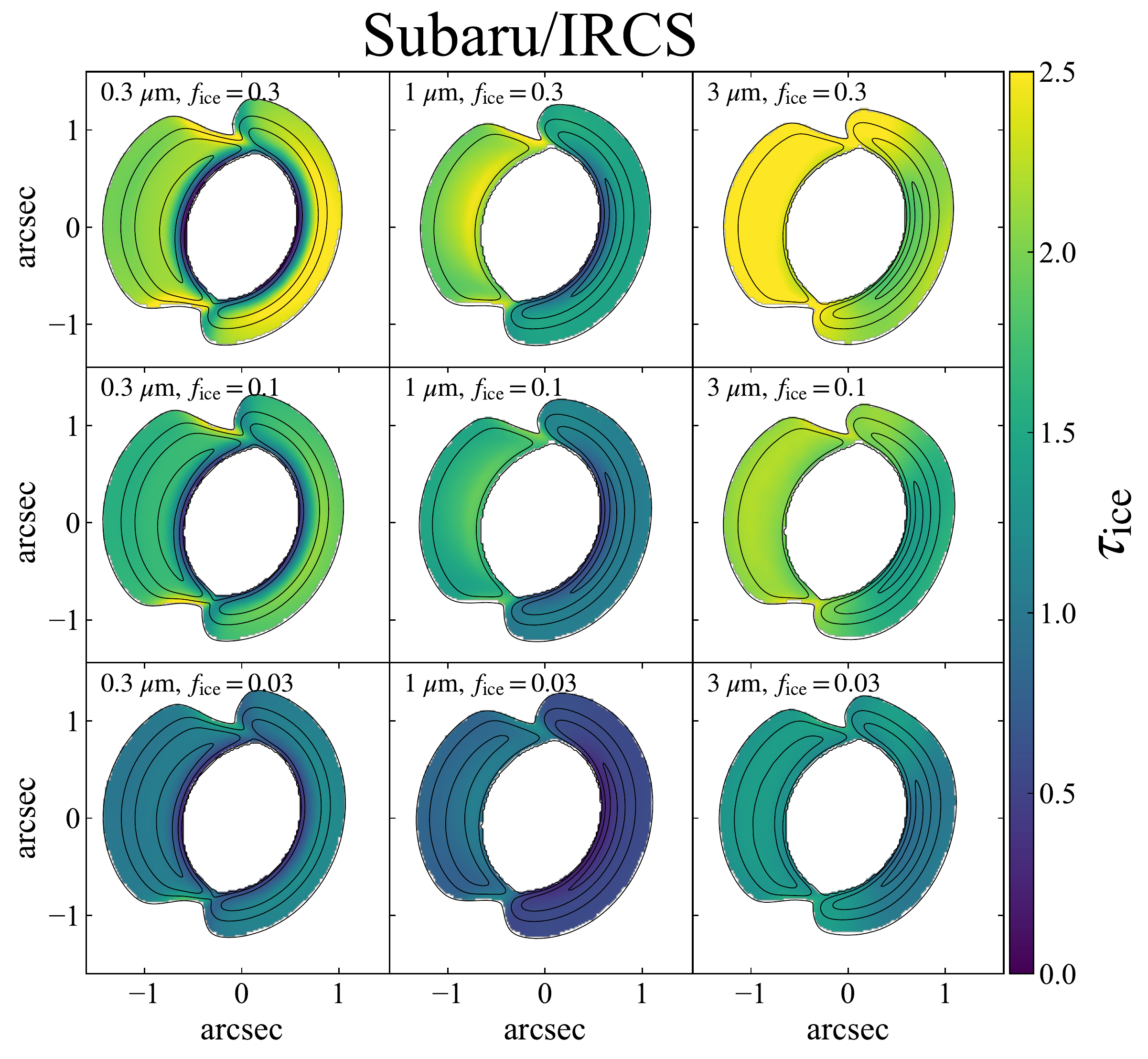}
\includegraphics[width=0.49\linewidth]{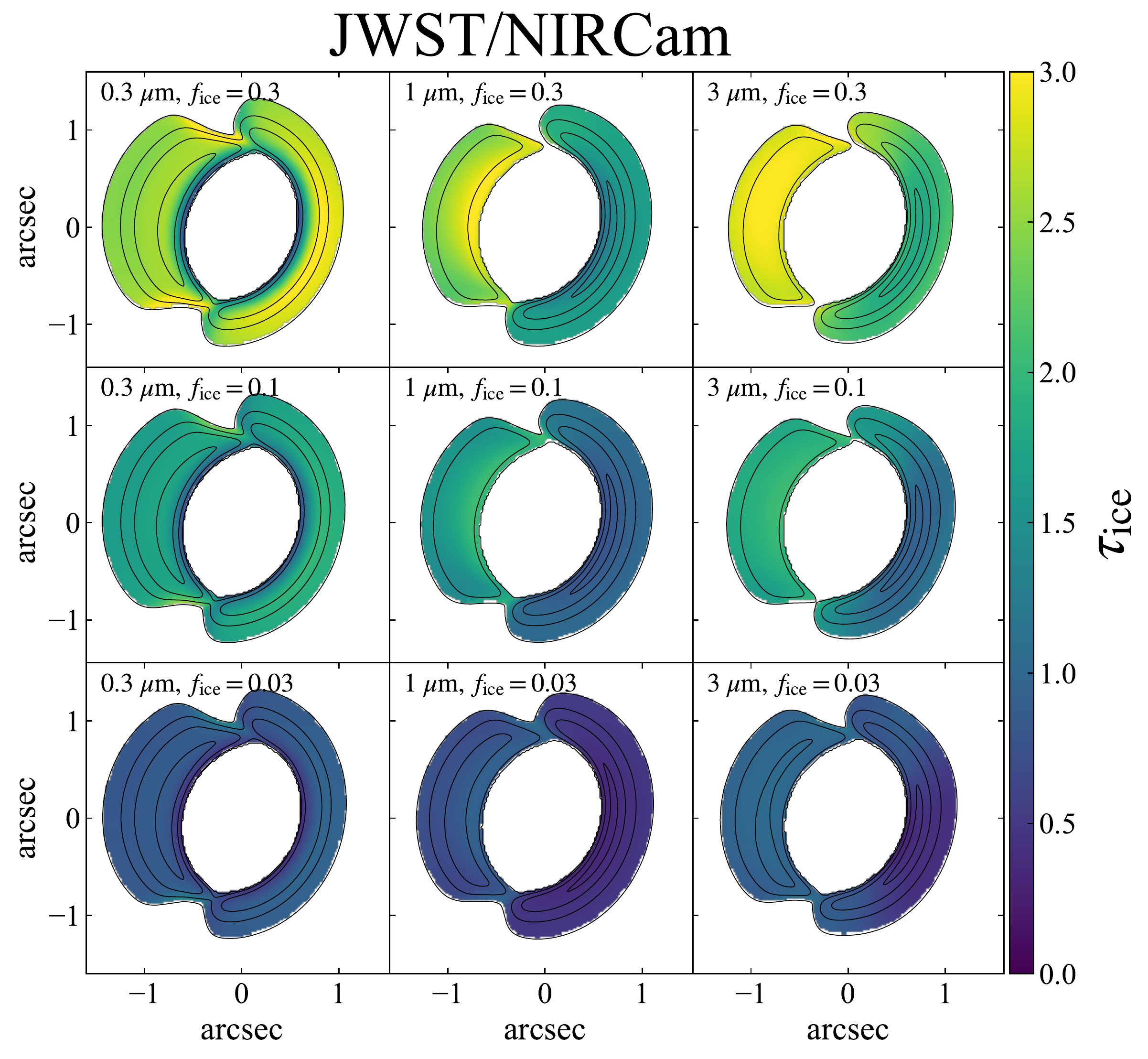}
\caption{The ice feature depth $\tauice$ for the outer disk for the Subaru/IRCS filters (left) and the JWST/NIRCam filters (right), respectively. The log-normal size distribution is assumed. For reference, the black contours show total intensity (3\%, 10\%, 30\%, 80\% of the peak intensity of the outer disk) at $\water$ band (left) and F300M (right) at each model. }
\label{fig:tauice}
\end{center}
\end{figure*}

Next, we investigate how the feature depth varies across the outer disk regions. We introduce {\it the effective optical depth} $\tauice$ as a feature depth measure. Since the scattering feature is an albedo effect rather than extinction along the line of sight, the feature depth does not reflect the actual optical depth.
However, it is still convenient to express the feature depth similarly to extinction features such that $\tauice$ is defined by \citep{Honda16}
\begin{eqnarray}
\tauice=\ln\left(\frac{I_\mathrm{cont}}{I_\mathrm{H2O}}\right), \label{eq:tauice1}
\end{eqnarray}
where $I_\mathrm{cont}$ is the continuum intensity at the $\water$ band estimated by 
\begin{eqnarray}
\log(I_\mathrm{cont})&=&\frac{\log(I_{L'})-\log(I_{K})}{\log\lambda_{L'}-\log\lambda_{K}}\nonumber\\
&&\times(\log\lambda_{\mathrm{ice}}-\log\lambda_{K})+\log(I_{K}),\label{eq:tauice2}
\end{eqnarray}
where $I_K$,  $I_\mathrm{H2O}$,  $I_{L'}$ and $\lambda_K$,  $\lambda_\mathrm{ice}$,  $\lambda_{L'}$ are total intensities and wavelengths at the $K$, $\water$, and $L'$ bands of Subaru/IRCS, respectively. Likewise, we calculate $\tauice$ for JWST/NIRCam filters, where we use total intensities at the F250M, F300M, and F360M filters instead of using those at the $K$, $\water$, and $L'$ bands in Equation (\ref{eq:tauice1} and \ref{eq:tauice2}). 

Figure \ref{fig:tauice} shows maps of $\tauice$ values for the outer disk with various grain radii and ice abundances. It turns out that even if the ice abundance is the same throughout the disk, the feature depth $\tauice$ is not uniform. In particular, for $a_c=3~\mu$m, the feature depth is asymmetric: the near-side of the disk tends to exhibit shallower features than the far-side. This is due to the wavelength dependence of phase function (Figure \ref{fig:phase}).

Figure \ref{fig:tauice} also demonstrates the feasibility of observing the scattering feature with JWST/NIRCam. Although the $\tauice$ values for the NIRCam filters are slightly different from those of the IRCS filters owing to the different observing wavelengths, the NIRCam filters (F250M, F300M, F360M) are sufficient to capture the ice feature. For example, $\fice=0.03-0.1$, the ice feature typically has a depth of $\tauice\sim0.5-2$. The near and far side asymmetry of the ice feature can also be seen with the JWST filters. 

Therefore, NIRCam coronagraphic imaging is a viable tool for investigating water ice for nearly face-on disks. Combined with NIRSpec spectroscopy for edge-on disks \citep[e.g.,][]{Ballering21}, JWST will reveal ice in disks at various inclination angles. However, careful analysis would be needed when estimating the amount of ice from scattering features with NIRCam, as the feature depth depends on the scattering properties of icy grains and disk geometry.

\section{Polarimetric spectra of disk-scattered light} \label{sec:result2}

The polarimetric spectra of the ice feature may provide another opportunity to constrain ice properties \citep{Pendleton90, Kim19, Tazaki21}. 
Although polarimetric spectra of the ice feature have not yet been reported in protoplanetary disks, this section aims to provide a model prediction for it.

\subsection{Scattering polarization feature of water ice} \label{sec:scatpol}

When light is scattered by a dust grain, it would be linearly polarized. This means that the scattering feature of ice can be polarized as well. Indeed, the degree of linear polarization of light scattered by large icy grains exhibits characteristic wavelength dependence across the 3 $\mu$m ice feature, which is termed as {\it the polarization feature} of water ice in \citet{Tazaki21}. 

For small grains, the polarization feature is less prominent as the scattering process obeys Rayleigh scattering. The polarization feature usually develops for grains larger than $\lambda/2\pi\sim0.5~\mu$m at $\lambda=3~\mu$m. For such large grains, the degree of linear polarization tends to be enhanced at $\lambda\sim3~\mu$m. The enhanced polarization degree is due to either the phase effect \citep{Pendleton90} or surface scattering of grains \citep{Tazaki21}.
 
\subsection{Polarization spectra for the outer disk of HD 142527}

To investigate what the polarization feature looks like in the outer disk of HD 142527, we calculate the polarization fraction of disk-scattered light spectra in our radiative transfer simulations.

Although, in Section \ref{sec:sizedist}, we found that a log-normal distribution is more favorable to explain the observed spectra, we here assume a power-law distribution model. This is because the log-normal size distribution causes unrealistic oscillatory behavior in polarization patterns originating from the perfect sphere assumption in the Mie theory. To smear out such oscillation, we adopt a power-law distribution.

\begin{figure}[t]
\begin{center}
\includegraphics[width=\linewidth]{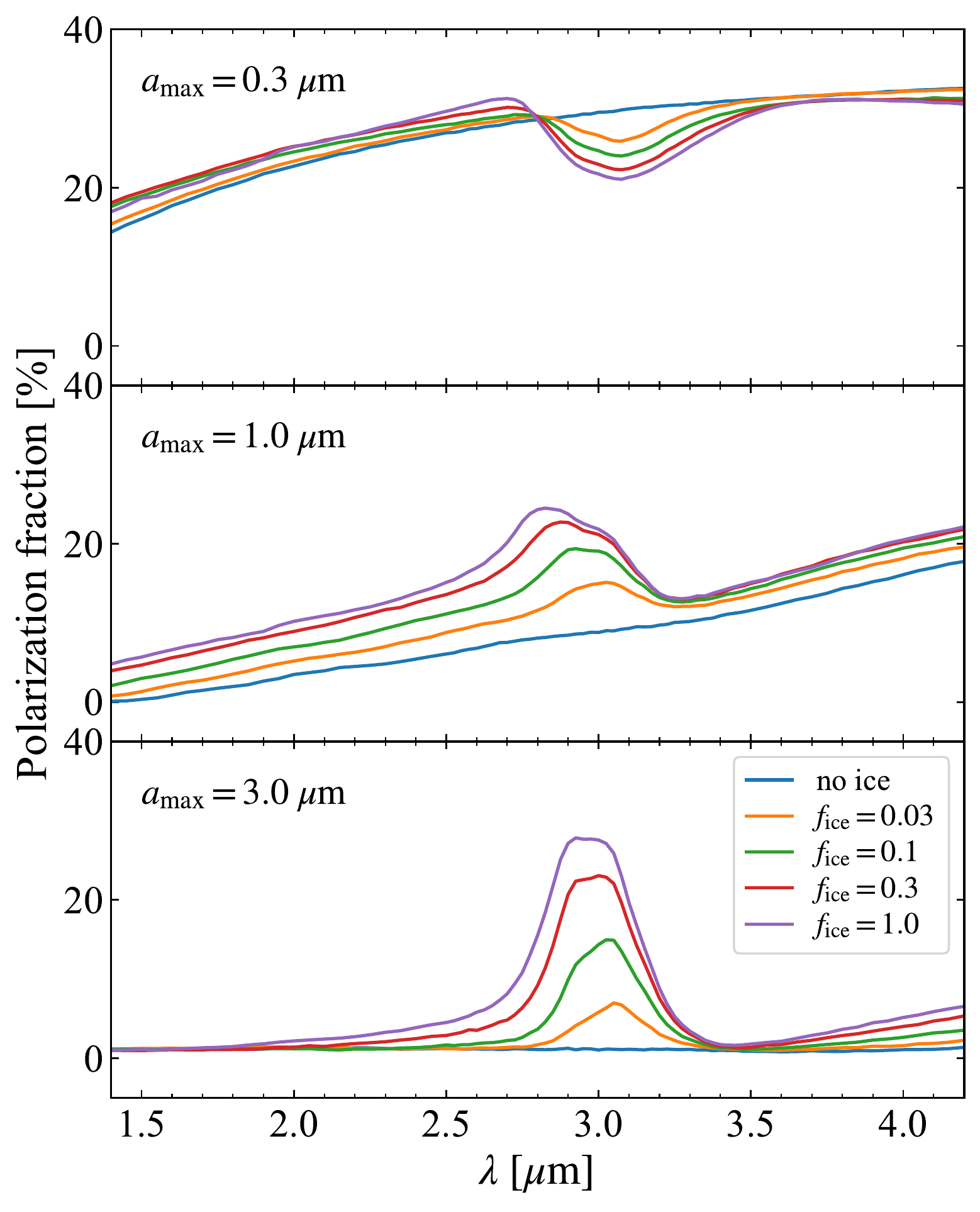}
\caption{The polarization fraction of the outer disk-scattered light at the near side (PA=$250^\circ$). The top, middle, and bottom panels show the results for $\amax=0.3~\mu$m, $1~\mu$m, and $3~\mu$m, respectively. }
\label{fig:polspec}
\end{center}
\end{figure}

Figure \ref{fig:polspec} shows polarization fractions measured at the near side of the outer disk (PA=$250^\circ$). 
When $\amax=0.3~\mu$m, grains approximately obey Rayleigh scattering; therefore, we do not expect the polarization feature \citep[see Figure 2 in][]{Tazaki21}. However, there is a slight reduction of polarization fraction at $\lambda\sim3~\mu$m. This is because, at $\lambda\sim3~\mu$m, the disk scattering surface comes at a higher altitude due to a larger extinction cross-section, which in turn decreases scattering angles at the near side. As a result, the polarization fraction is slightly reduced at this wavelength.

Once grain radius exceeds $\sim0.5~\mu$m, the polarization feature starts to develop, and consequently, the polarization fraction is enhanced at $\lambda\sim3~\mu$m. The polarization excess depends on an ice abundance, and a higher ice abundance gives rise to a larger excess in polarization fraction. 

We measure the excess of the polarization fraction by $\delp=P-\pcont$, where $\pcont$ is the estimated continuum polarization fraction:
\begin{equation}
\pcont=\frac{P_{L'}-P_{K}}{\lambda_{L'}-\lambda_{K}}(\lambda_{\water}-\lambda_{K})+P_{K},
\end{equation}
where $P_{K}$, $P_{\water}$, and $P_{L'}$ are polarization fractions at the $K$ band, the $\water$ band, and the $L'$ band, respectively. 

\begin{figure}[t]
\begin{center}
\includegraphics[width=\linewidth]{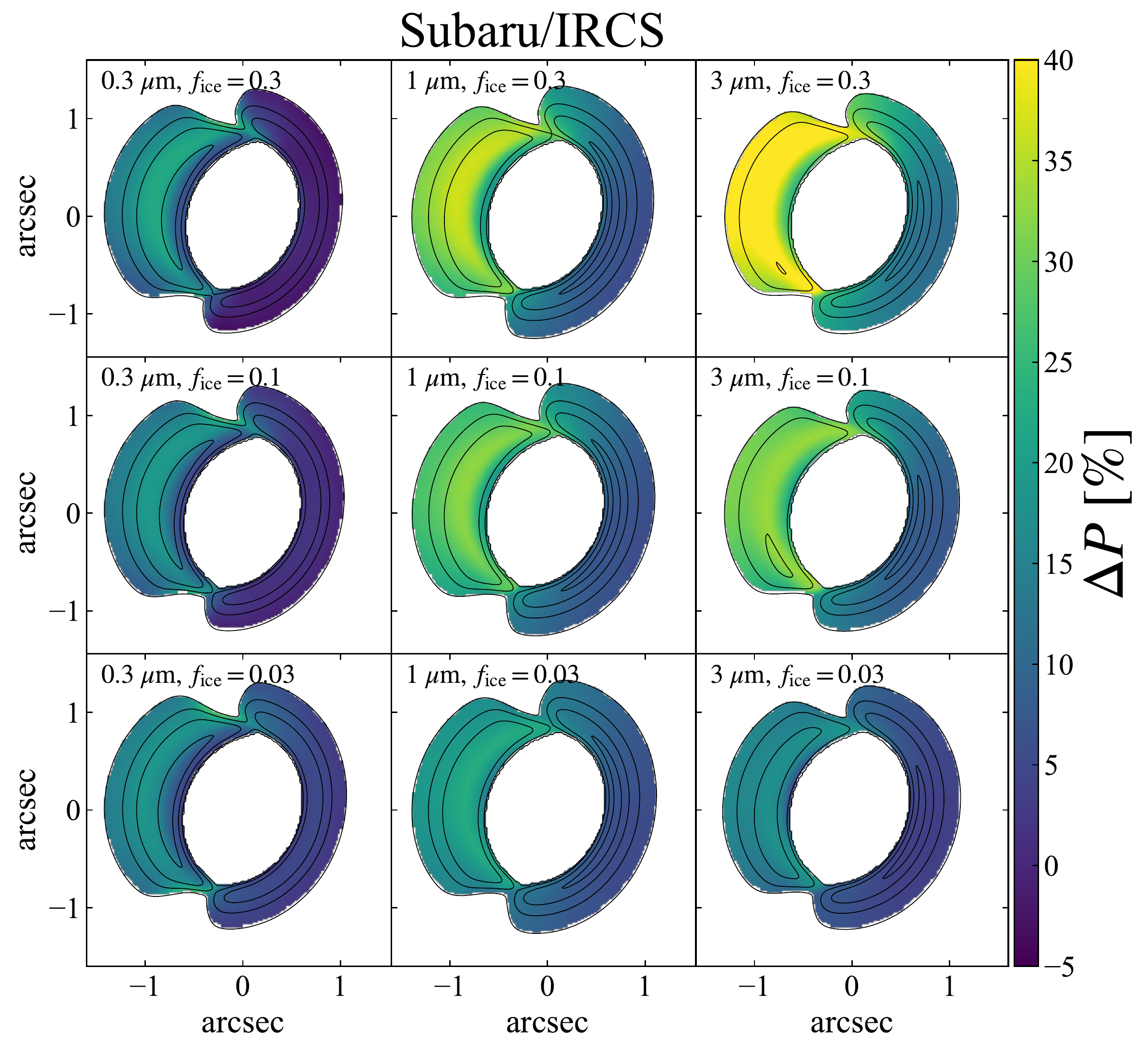}
\caption{Maps of the polarization excess $\delp$ at the $3~\mu$m ice feature for the outer disk.}
\label{fig:polmap}
\end{center}
\end{figure}

Figure \ref{fig:polmap} shows maps of $\delp$ values for the outer disk. The polarization excess $\delp$ shows a near-and-far side asymmetry for all grain radii and ice abundances, and the far side of the disk usually exhibits a larger value of $\delp$ than the near side. When $\amax=3~\mu$m and $\fice=0.03$, the polarization excess would be $\sim5\%$ for the near side and $\sim15\%$ for the far side. 

Polarization fractions of the outer disk of HD 142527 have been measured at near-IR wavelengths. \citet{Canovas13} found that $H$-band polarization fractions vary $10-25\%$ across the outer disk, while \citet{Avenhaus14} obtained higher values $20-45\%$. Most recently, \citet{Hunziker21} carried out precise measurements of polarization fraction at VBB and $H$ bands and found that the polarization fraction at the near side is around $20.2\pm0.6$ \% at the $H$ band. These values are significantly higher than those predicted in our models for both $\amax=1~\mu$m and $3~\mu$m. 

One possibility to reconcile this difference is to consider porous dust aggregates. Light scattering simulations have shown that micron-sized aggregates exhibit a significantly higher polarization fraction \citep{Min16, Tazaki16}. 
Although the porosity should not be extremely high to explain the observed reddish spectra \citep{Tazaki19a}, we anticipate moderately porous aggregates may be responsible for both reddish spectra and a high polarization fraction. To draw a robust conclusion, light scattering simulations for such dust aggregates are necessary, although this is beyond the scope of this paper.

\subsection{Can we see the ice feature in polarized intensity?}

As near-IR disk observations have often employ polarization differential imaging techniques to suppress a bright central stellar light, it would be interesting to see what the ice feature looks like in polarized intensity.

\begin{figure}[t]
\begin{center}
\includegraphics[width=\linewidth]{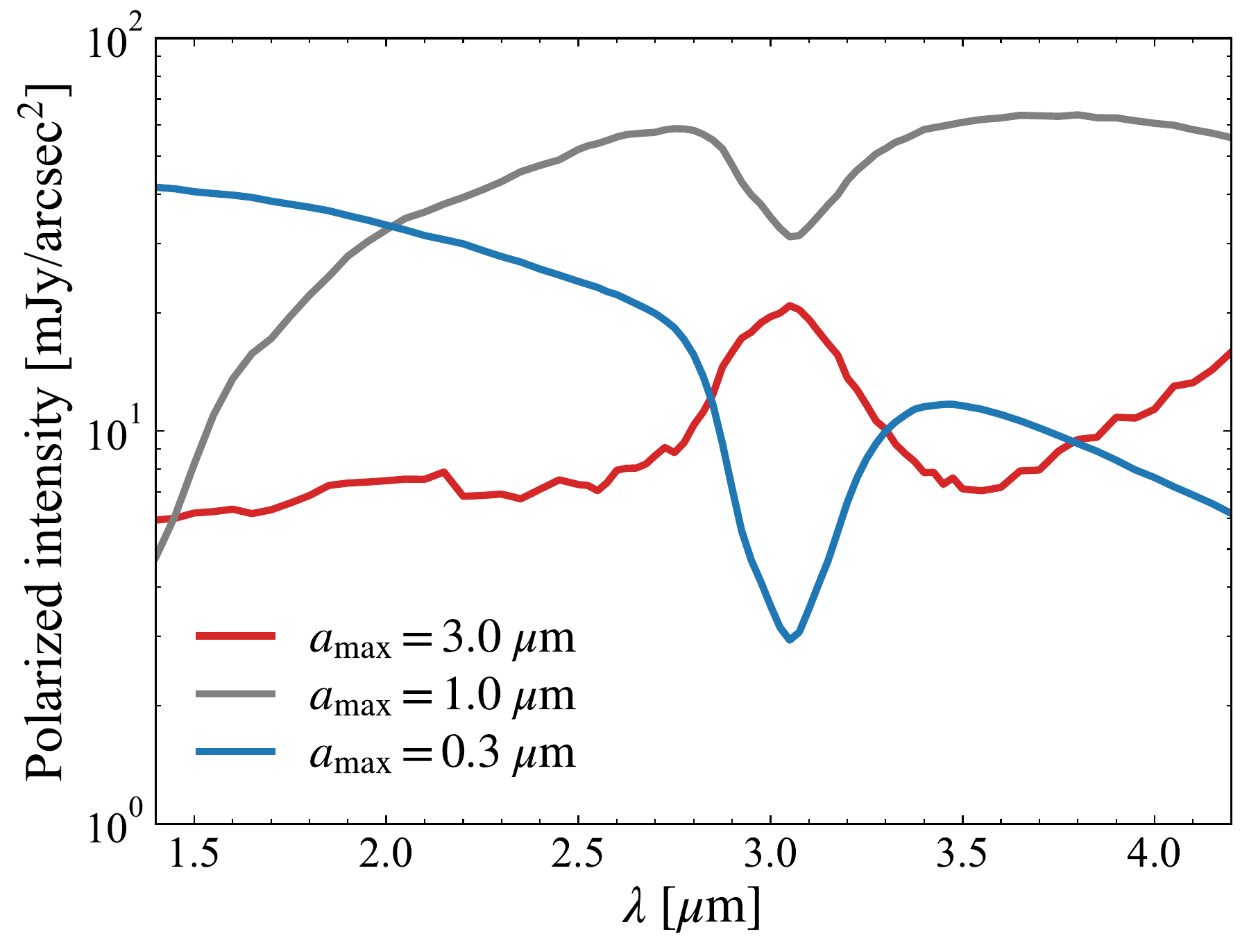}
\caption{The model spectra of polarized intensity measured at the near side of the disk (PA=250$^\circ$). The blue, gray, and red lines are the results for $\amax=0.3~\mu$m, 1~$\mu$m, and $3~\mu$m, respectively. The ice abundance is assumed to be $\fice=0.03$. The small scale jitter in the red line is photon noise in our Monte Carlo radiative transfer simulations.}
\label{fig:pispec}
\end{center}
\end{figure}

Figure \ref{fig:pispec} shows the polarized intensity spectra for three different grain radii: $\amax=0.3$, 1, 3 $\mu$m with $\fice=0.03$. Similar to the case of total intensity, we can clearly see the ice feature in polarized intensity for small grains ($\amax=0.3~\mu$m, 1 $\mu$m), where the spectra show a dip at $\lambda\sim3~\mu$m. In contrast, for the case of $\amax=3~\mu$m, the polarized intensity exhibits an excess at $\lambda\sim3~\mu$m as the polarization fraction is enhanced at this wavelength (see also Figure \ref{fig:polspec}). Thus, the ice feature may be also identified in polarized intensity spectra. 

However, the lack of the ice feature in polarized intensity does not necessarily mean the absence of ice. Even if ice exists in the disk surface layer, the ice feature may not exhibit a distinct peak in the polarized intensity when the absorption in total intensity and the excess of polarization fraction counteract each other. 

\section{Discussion} \label{sec:disc}

\subsection{Water-ice abundance at the disk surface} \label{sec:ice}

We discuss water ice in disks around HD 142527 and HD 100546.

\subsubsection{HD 142527}
The ice/silicate mass ratio inferred by our model is approximately $0.06-0.2$, which is significantly smaller than that inferred by \citet{Honda09}, where they estimated the ratio of 2.2 or even higher. As shown in Section \ref{sec:sizedist}, this difference is due to the ignorance of anisotropic scattering in the previous study \citep{Inoue08}. Anisotropic scattering has a role in making the feature deeper, and consequently, our model comes at a relatively low ice abundance compared to the previous study.

The outer disk of HD 142527 is known to exhibit ice emission features in far-IR wavelengths \citep{Min16}. Based on the SED modeling, \citet{Min16} derived the ice/silicate mass ratio of $1.6^{+0.9}_{-0.6}$. This value is much larger than our estimate. However, far-IR emission features are arising from cold ice ($\sim 45$ K), whereas the near-IR scattered light traces warm ice ($\gtrsim70$ K) (see also Appendix \ref{sec:amo}). Hence, the mismatch in ice abundance might reflect spatial variation of ice abundance, i.e., a lower ice abundance for an upper disk surface. 

A promising mechanism to reduce water ice at a higher disk surface is photodesorption caused by far-ultraviolet photons \citep{Dominik05, Oka12, Furuya13, Kamp18, Ballering21}. \citet{Oka12} examined the impact of photodesorption on the location of the water snowline at the disk surfaces of HD 142527, and they found that the snowline is likely located between 100 au and 300 au, which is similar to the region seen in scattered light in \citet{Honda09}. Thus, it seems possible that the observed region is close to the water snowline, and then the inferred ice abundance becomes very low.

An alternative mechanism is ro-thermal desorption of water ice \citep{Hoang20, Tung20}. The stellar radiation field can drive a rapid spin for grains via radiative torques, and this results in either rotational disruption of grains and/or ro-thermal desorption of ice mantles \citep{Hoang20}. 
Once the rotational disruption operates, the centrifugal force tears apart micron-sized grains into small fragments, producing a bottom-heavy grain-size distribution. Since scattered light spectra favor depletion of such small grains (Section \ref{sec:sizedist}), the role of the rotational disruption at the scattering surface of the outer disk of HD 142527 seems limited. Nevertheless, a moderately rapid spin can help water molecules to desorb from the ice mantle via ro-thermal desorption. 

A more detailed study of photodesorption and ro-thermal desorption at the HD 142527 disk would be helpful to clarify which mechanism is responsible for the inferred low ice abundance.

\subsubsection{HD 100546}

The disk around HD 100546 is another object exhibiting the scattering feature \citep{Honda16}. Interestingly, this object is also known to show reddish scattered light from optical to near-IR wavelengths \citep{Mulders13, Sissa18}, and $2.5~\mu$m-sized grains seem to be lifted above the scattering surface \citep{Mulders13}. Hence, we suppose the grain-size distribution is similar to the log-normal distribution, as with the case of the outer disk of HD 142527.

\citet{Honda16} detected the scattering feature of ice and compared the feature depth at different locations of the disk. Along the major axis, they found $\tau_\mathrm{ice}$ increases approximately from $\sim0.5$ to $\sim1.0$ for larger radial distances. These values are compatible with our results when $\fice=0.03$ and $a_c=1-3~\mu$m (Figure \ref{fig:tauice}). Thus, grains in the HD 100546 disk might be ice-poor as well. 

\citet{Honda16} also found that $\tau_\mathrm{ice}$ values are asymmetric along the minor axis: larger $\tau_\mathrm{ice}$ for the near side. This asymmetry is, however, the opposite to the one anticipated from our simulations (Figure \ref{fig:tauice}), which predicts larger $\tau_\mathrm{ice}$ for the far side. We thus speculate that the observed asymmetry reflects the variation of ice abundance. This seems plausible because the near and far sides of the disk trace different radial distances from the central star owing to disk flaring \citep{Stolker16}. In this case, the asymmetry can arise from the radial variation of ice abundance, as seen along the major axis.

\subsection{Vertical grain dynamics in the outer disk of HD 142527}\label{sec:alpha}

We showed in Section \ref{sec:spec} that micron-sized grains are necessary to explain the observed reddish spectra. Because such large grains are usually expected to settle down below the scattering surface, our results require a mechanism that hinders dust settling. As such, we consider turbulent diffusion of grains \citep{Dub95, Dullemond04, Youdin07} and discuss how large a diffusion coefficient would be needed to explain the observations.

\subsubsection{Estimation of dust diffusion coefficients}

We compare two timescales relevant to vertical grain dynamics to estimate the diffusion coefficient: vertical settling and stirring \citep[e.g.,][]{Dullemond04}. 
The stirring timescale is given by $t_\mathrm{stir}=z^2/D$, where $z$ is the height from the midplane, and $D=D_0/\mathrm{Sc}$ is the diffusion coefficient of dust grains in the vertical direction; $\mathrm{Sc}$ is the Schmidt number, $D_0=\alpha c_sH_g$ is the diffusion coefficient of gas molecules, $\alpha$ is a non-dimensional parameter, $c_s$ is the sound speed, and $H_g$ is the pressure scale height. We assume $\mathrm{Sc}=1$ in the following. 

The settling timescale is given by $t_\mathrm{sett}=z/v_\mathrm{sett}$, where $v_\mathrm{sett}$ is the settling velocity. In the terminal velocity approximation, the settling velocity $v_\mathrm{sett}$ is given by the force balance between gas drag and gravity in the $z$ direction: 
\begin{eqnarray}
\frac{4}{3}\rho_\mathrm{g}\left(\frac{\sigma}{m}\right) v_\mathrm{th} v_\mathrm{sett}f_d=\Omega^2z, \label{eq:fz}
\end{eqnarray}
where $\rho_\mathrm{g}$ is the gas density, $\sigma/m$ is the area-to-mass ratio of the dust particle, $v_\mathrm{th}=\sqrt{8/\pi}c_s$ is the thermal velocity of gas molecules, $\Omega=c_s/H_g$ is the Kepler angular frequency, and 
\begin{equation}
f_d=\sqrt{1+\frac{9\pi}{128}\left(\frac{v_\mathrm{sett}}{c_s}\right)^2},
\end{equation}
is a simplified correction factor describing the drag force in the supersonic regime \citep{Kwok75, Paardekooper07}.
By solving Equation (\ref{eq:fz}), the settling velocity can be expressed as
\begin{equation}
\left(\frac{v_\mathrm{sett}}{c_s}\right)^2=\frac{64}{9\pi}\left[\sqrt{1+\left(\frac{9\pi}{64}\frac{m}{\sigma}\frac{z}{\rho_\mathrm{g}H_\mathrm{g}^2}\right)^2}-1\right]. \label{eq:vsett}
\end{equation}

In order for grains to be stirred up to the surface layer, we anticipate $t_\mathrm{stir}\lesssim\xi t_\mathrm{sett}$ at the surface, which yields
\begin{eqnarray}
\alpha&\gtrsim& \xi^{-1}\left(\frac{z}{H_\mathrm{g}}\right)\left(\frac{v_\mathrm{sett}}{c_s}\right),\\
&=&3\times10^{-1}\left(\frac{\xi}{10}\right)^{-1}\left(\frac{z}{3H_\mathrm{g}}\right)\left(\frac{v_\mathrm{sett}}{c_s}\right).
\end{eqnarray}
\citet{Dullemond04} estimated that $\xi=100$ provides a good indication of a depletion height of dust grains. Because $\xi=100$ would be too optimistic to estimate $\alpha$, we here adopt $\xi=10$.
We also assume that the scattering surface is at $z=3H_g$, which is approximately consistent with our radiative transfer model at $\lambda\sim3~\mu$m. 

To estimate the gas density in Equation (\ref{eq:vsett}), we assume the hydrostatic equilibrium in the vertical direction with the gas surface density $\Sigma_\mathrm{g}$. Based on CO observations with Atacama Large Millimeter/submillimeter Array, \citet{Boehler17} estimated $\Sigma_\mathrm{g}\simeq0.3$ g cm$^{-2}$ at a southern region of the outer disk of HD 142527. A similar gas surface density has been inferred as well in other studies \citep{Muto15, Garg21}. It is worth bearing in mind however that disk gas mass derived from CO emission can be substantially underestimated (see Section \ref{sec:gasmass}).

The area-to-mass ratio of dust particles is also needed to determine the settling velocity. We first consider spherical grains, and later on, porous dust aggregates (Section \ref{sec:por}). For spherical grains, $\sigma/m=3/(4\rho_\mathrm{m}a)$, where $\rho_\mathrm{m}$ is the material density. In the limit of the subsonic regime ($f_d\to1$), we except $\alpha\propto(m/\sigma)\propto a$. 

\begin{figure}[t]
\begin{center}
\includegraphics[width=\linewidth]{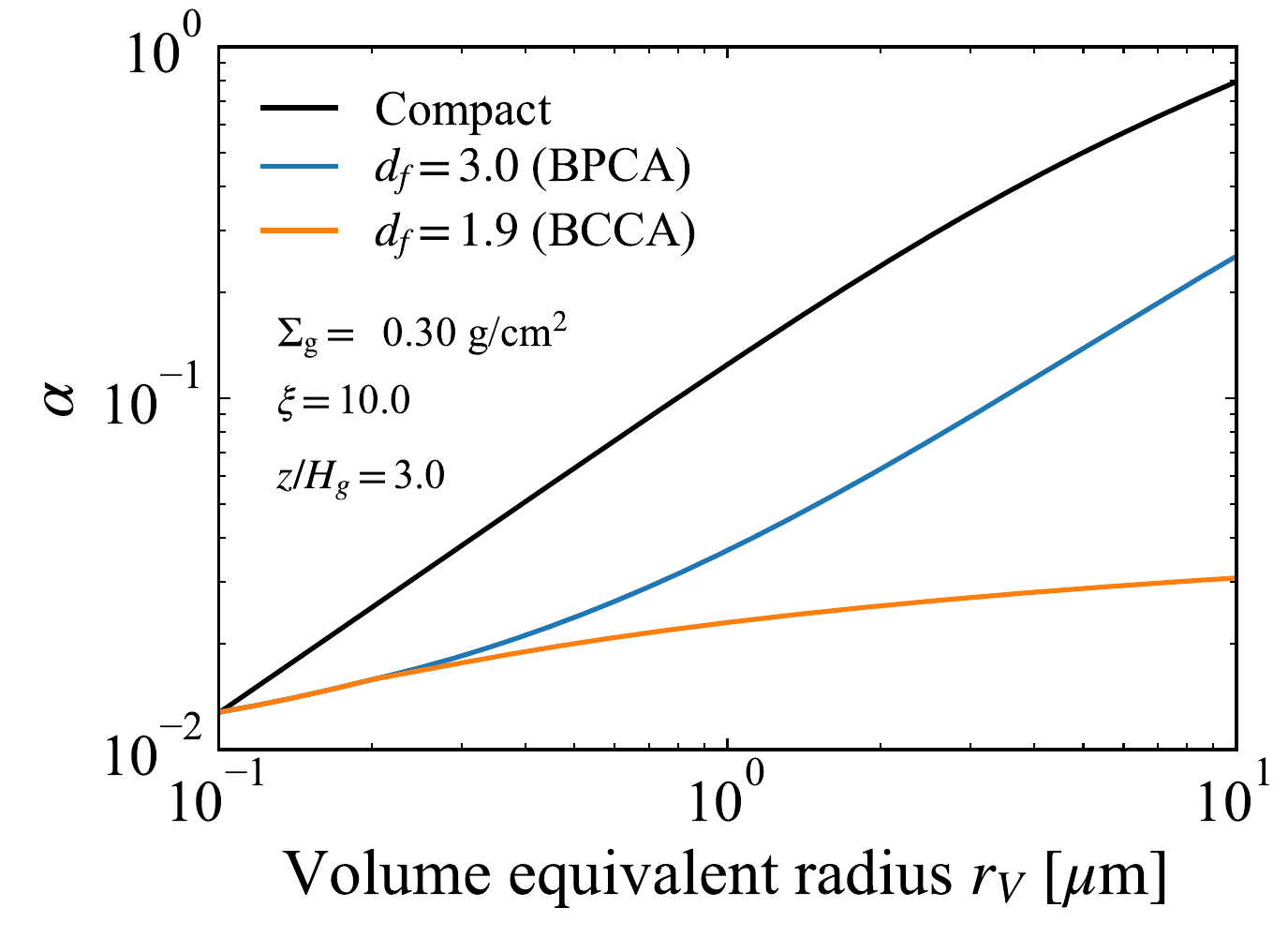}
\caption{The minimum value of $\alpha$ needed to keep spherical dust grains or dust aggregates above the scattering surface. The black, blue, and orange lines represent the results for compact spheres, BPCA and BCCA clusters, respectively. For porous aggregates, the monomer radius is assumed to be 0.1  $\mu$m.}
\label{fig:alpha}
\end{center}
\end{figure}

Figure \ref{fig:alpha} shows the minimum value of $\alpha$ required for the turbulence to oppose dust settling. A higher $\alpha$ value is necessary for larger grains owing to their worse dynamical coupling. Since the grain radius should be about 3 $\mu$m (Figure \ref{fig:spec}), we need $\alpha\gtrsim0.3$ to lift them above the surface. The inferred value is similar to the results in \citet{Mulders13}, where they derived $\alpha\gtrsim0.2$ to explain the reddish disk-scattered light of HD 100546.

The inferred value of $\alpha$ is surprisingly high compared to values inferred by gas observations for other disks. Molecular line observations have often resulted in non-detections of non-thermal motion of gas molecules and have placed an upper limit $\alpha<(3-10)\times10^{-3}$ at the disk surfaces \citep{Flaherty15, Teague16, Flaherty17, Flaherty18, Teague18}. Although recent observations have detected a prominent non-thermal motion for some disks, i.e., $\alpha=0.078\pm0.002$ for DM Tau \citep{Flaherty20} and $\alpha\sim0.05$ for HD 135344 B \citep{Casassus21}, these values are still below our inferred value.

There are two possibilities to mitigate this issue: (1) the gas density is currently underestimated, and (2) the area-to-mass ratio of dust particles is underestimated. We discuss each possibility below and summarize it in Section \ref{sec:alphasum}.

\subsubsection{Higher gas surface density?} \label{sec:gasmass}

The current estimates of the gas disk mass of HD 142527 hinge on CO observations \citep{Muto15, Boehler17, Garg21}. However, recent studies have shown that CO emission lines are not always be a reliable gas mass tracer \citep[e.g.,][]{Bergin17}. HD line observations with Herschel \citep{Bergin13, McClure16} have revealed that CO molecules are substantially depleted from disk surfaces compared to the ISM abundance \citep{Favre13, Nomura16, Schwarz16, McClure16}. Such depletion is likely due to both chemical processing of CO into less volatiles molecules and physical sequestration of CO in the form of ice \citep[][references therein]{Krijt20}.

\citet{Kama20} revisited HD emission observation data of Herschel and updated disk mass constraints for 15 disks, including HD 142527. However, the gas disk mass for HD 142527 is only loosely bounded, and a gravitationally unstable disk is still allowed within the limits of the Herschel HD data. 
Because the degree of CO depletion in HD 142527 remains uncertain, we cannot exclude the possibility that the gas disk mass is significantly higher than the current estimate.

We thus consider an extreme case, where the observed spiral structures \citep[e.g.,][]{Avenhaus14} are attributed to gravitational instability, and then we assume $Q=1.5$, where $Q$ is the Toomre's $Q$ value. We also assume an axisymmetric gas disk with a temperature of 45 K. As a result, $Q=1.5$ yields $\Sigma_\mathrm{g}=12.8$ g cm$^{-2}$ at a radial distance of 205 au, which corresponds to the peak location of the southern gas density \citep{Boehler17}. The derived surface density is approximately 43 times higher than the current estimate ($\Sigma_\mathrm{g}=0.3$ g~cm$^{-2}$), and therefore the total gas disk mass should be around 0.24$M_\sun$. Because the dust mass inferred by \citet{Boehler17} is $1.5\times10^{-3}M_\sun$, our massive disk model has a gas-to-dust ratio of around 160, which is similar to the canonical interstellar value of 100. Thus, our massive disk model has a reasonable gas-to-dust ratio, despite the significant increase in gas disk mass.

This massive disk model provides $\alpha\gtrsim9\times10^{-3}$ for $3 ~\mu$m spherical grains, which is lower than the nominal value by a factor of 38. Therefore, the massive disk model can significantly relax the requirement for $\alpha$ values.

\subsubsection{Highly porous dust aggregates?} \label{sec:por}

An alternative solution to lower $\alpha$ values is considering porous aggregates, which exhibit a large area-to-mass ratio and strong dynamical coupling with gas. Porous aggregates are favorable to explain the observed high polarization fractions at a near-IR wavelength \citep{Canovas13, Avenhaus14, Hunziker21}, as large dust aggregates tend to produce a highly polarized scattered light \citep[e.g.,][]{Min16, Tazaki16}.

To examine the effect of porosity on $\alpha$, we consider two models for porous dust aggregates \citep[e.g., see Figure 1 in][]{Tazaki19a}: ballistic cluster-cluster aggregation (BCCA) and ballistic particle-cluster aggregation (BPCA). BCCA and BPCA clusters have fractal dimensions of 1.9 and 3.0 and porosities of $>90\%$ and $85\%$, respectively \citep[e.g.,][]{Kozasa92}. The area-to-mass ratio of these aggregates can be computed analytically with a technique recently developed in \citet{Tazaki21b} \footnote{The codes are available at the author's GitHub repository \url{https://github.com/rtazaki1205/geofractal}}.

Figure \ref{fig:alpha} shows the necessary values of $\alpha$ for BCCA and BPCA clusters. To specify the size of aggregates, we use the volume equivalent radius, $r_V=r_0N^{1/3}$, where $r_0$ is the monomer radius, and $N$ is the number of monomers. We assume $r_0=0.1~\mu$m. 

As expected, the necessary values of $\alpha$ are smaller for aggregates with higher porosity. For BPCA and BCCA clusters at $r_V=3.0~\mu$m, the $\alpha$ values needed are reduced by a factor of 3.8 and 12.5 compared to the spherical grains, respectively.

However, \citet{Tazaki19a} pointed out BCCA clusters with 0.1 $\mu$m-sized monomers are unlikely to cause reddish scattered lights at near-IR wavelengths (see, e.g., Figures 5 and 10(left) in their paper). Thus, we rule out BCCA clusters and assume BPCA clusters as aggregates in the disk. In this case, we obtain $\alpha\gtrsim9\times10^{-2}$ at $r_V=3~\mu$m. 

\subsubsection{Summary of the required turbulent strengths} \label{sec:alphasum}
\begin{figure}[t]
\begin{center}
\includegraphics[width=\linewidth]{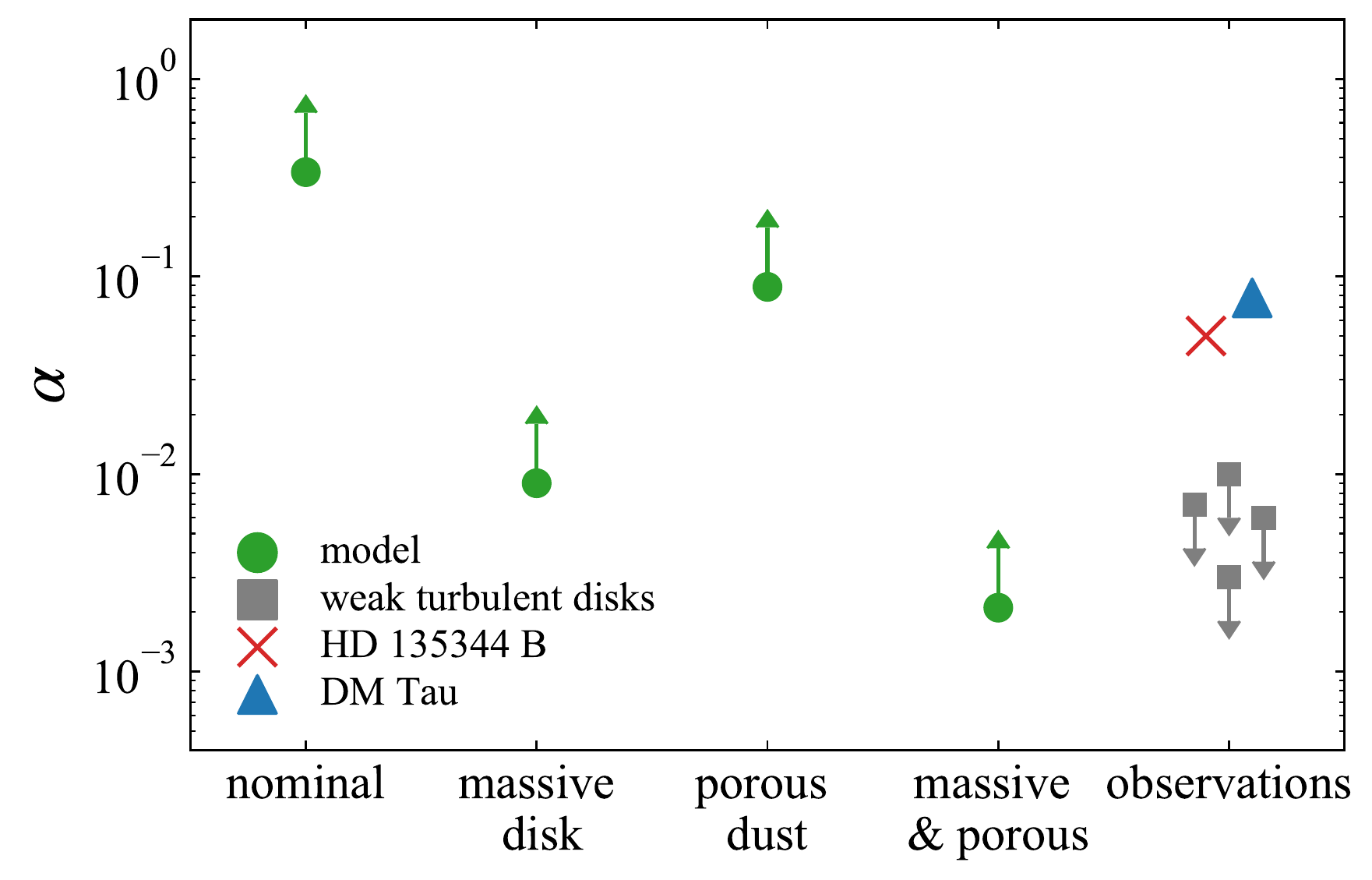}
\caption{The minimal values of the $\alpha$ parameter for the vertical diffusion coefficient for various disk and dust models: nominal ($\Sigma_\mathrm{g}=0.3$ g cm$^{-2}$, spherical grains), massive disk ($\Sigma_\mathrm{g}=12.8$ g cm$^{-2}$, spherical grains), porous dust ($\Sigma_\mathrm{g}=0.3$ g cm$^{-2}$, BPCA clusters), and massive \& porous ($\Sigma_\mathrm{g}=12.8$ g cm$^{-2}$, BPCA clusters) models. The upper and lower arrows represent the lower and upper limits for $\alpha$ values, respectively. We assume the same diffusion coefficients for dust and gas ($\mathrm{Sc}=1$). The $\alpha$ values from line-broadning observations are taken from \citet{Flaherty20} (DM Tau), \citet{Casassus21} (HD 135344 B), \citet{Flaherty17,Flaherty18,Flaherty20} (upper limits for TW Hya, HD 163296, V4046 Sgr, MWC 480).}
\label{fig:alphamodel}
\end{center}
\end{figure}

Figure \ref{fig:alphamodel} summarizes the inferred values of $\alpha$ for several models. As already mentioned, in our nominal case (spherical grains and $\Sigma_\mathrm{g}=0.3$ g~cm$^{-2}$), we need $\alpha\gtrsim0.3$. 

For the massive disk model (spherical grains and $\Sigma_\mathrm{g}=12.8$ g~cm$^{-2}$),  we obtained $\alpha\gtrsim9\times10^{-3}$ and this is well below the observed `strong turbulence' level \citep{Flaherty20, Casassus21}, which is $\alpha\sim 0.05-0.08$. Meanwhile, the porous dust model relaxes the constraint only by a factor of 3.8, and hence, the impact seems limited. From these results, we conclude that the disk gas mass is the dominant factor to lower $\alpha$ values. Our conclusion implies that the current gas disk mass for HD 142527 derived from CO emission observations is likely underestimated unless the actual turbulence is extremely strong ($\alpha\gtrsim0.3$).

We can further reduce the lower limit down to $\alpha\gtrsim2\times10^{-3}$, once both massive disk and porous dust models are employed. This case is thought to provide the lowest $\alpha$ value. Nevertheless, the required value is still high, i.e., compared to the disk around HL tau in which $\alpha\sim10^{-4}$ has been suggested \citep{Pinte16}. Hence, vertical grain dynamics in the outer disk of HD 142527 is different from that of HL Tau, and a relatively efficient vertical mixing or dust lifting process should operate in the disk.

Since the lowest $\alpha$ value inferred is close to the current upper limits of molecular line observations \citep{Flaherty15, Teague16, Flaherty17, Flaherty18, Teague18}, we may expect a kinetic detection of such turbulent motion. Although the gas flow is not necessary to be turbulent (see Section \ref{sec:wind}), such observations would be helpful to reveal the origin of vertical grain dynamics.

\subsection{Origin of the dust lifting} \label{sec:lift}

In Section \ref{sec:alphasum}, we estimated that $\alpha\sim2\times10^{-3}$ is at least necessary to keep micron-sized grains at the disk surface of HD 142527. We then proceed to explore which mechanism may potentially explain the inferred `dust lifting' in the disk.

\subsubsection{Turbulent stirring}

The vertical shear instability (VSI) has recently attracted attention as a candidate for disk turbulence \citep{Stoll16, Flock17, Flock20, Lin19}. The VSI results in strong vertical mixing of grains, which occurs much more efficiently than transporting the angular momentum radially. \citet{Flock20} estimated $\alpha\sim5.4\times10^{-3}$ for the vertical diffusion coefficient of grains, and this is above our minimally required value. However, the $\alpha$ value would be lowered once disk magnetization \citep{Cui20, Cui21} or grain growth is considered \citep{Fukuhara21}. In addition, the VSI may not operate at disk surfaces, where slow thermal relaxation makes the VSI being stabilized by buoyancy forces \citep[e.g.,][]{Fukuhara21}. Thus, the VSI is unlikely to be the source of the inferred strong turbulence at the disk surface.

The required values of $\alpha\gtrsim2\times10^{-3}$ are reminiscent of vigorous turbulence seen in ideal MHD disks \citep{Fromang09, Zhu15}. In this case, the vertical diffusion coefficient increases for a higher altitude of disks \citep{Fromang09}. Although ambipolar diffusion tends to suppress MHD turbulence, the required vertical diffusion coefficient can be generated, particularly when a plasma $\beta$ is as low as $10^{3}$ \citep{Zhu15, Riols18} (see also \citet{Hasegawa17}). Therefore, MHD turbulence is a possibility for the inferred dust lifting.

Meanwhile, it is interesting to notice that reddish scattered light disks are often associated with spiral structures. For HD 142527, we can see spiral structures \citep[e.g.,][]{Hunziker21}. The following disks also exhibit both reddish spectra and spiral structures: HD 100546 \citep{Mulders13, Sissa18}, HD 135344 B (SAO 206462) \citep{Maire17}, and HD 100453 \citep{Long17}.
 
The spiral structure may cause efficient vertical diffusion of grains \citep[e.g.,][]{Bae16b, Bae16a}. If the observed spirals in HD 142527 are due to gravitational instability, gravoturbulence would also help to mix grains vertically as well \citep{Riols20}. Recently, \citet{Casassus21} detected a prominent non-thermal motion of CO molecules in the disk around HD 135344 B, which corresponds to $\alpha\sim0.05$, and argued that the motion seems to link its spiral structures. Therefore, turbulence within spiral structures is another fascinating possibility to explain our results.

It should be noted, however, that disks with spiral structures do not always give rise to reddish disk-scattered light. The following objects have been confirmed to have spiral structures, but they show blue scattering in near-IR wavelengths: LkH$\alpha$ 330 \citep{Uyama18}, HD 34700A \citep{Monnier19, Uyama20}, and AB Aur \citep{Fukagawa10, Hashimoto11, Boc20}. Recent studies have proposed that the spiral structures in the AB Aur disk might have formed by the late infall of material on the disk \citep{Dullemond19, Kuff20}. If this is the case, the disk, including the spiral structures, might be rich in smaller dust particles, thereby accounting for the blueish nature of these spiral disks.

\subsubsection{Disk winds} \label{sec:wind}

We have so far assumed that turbulence is responsible for opposing dust settling; however, a disk wind is another possibility. 

\citet{Miyake16} investigated the vertical distribution of grains under disk winds driven by the magnetorotational instability (MRI). They found that micron-sized grains accumulate at the disk surfaces. However, for the outer disk of HD 142527, the non-dimensional stopping time $T_s$ is estimated to be $\sim10^{-4}$ for $3~\mu$m grains at the disk midplane even if we adopt the massive disk model (see Section \ref{sec:gasmass}). This is larger than the accumulation condition $T_s\approx(4.5-8.1)\times10^{-6}$ found in \citet{Miyake16}; hence, micron-sized grains would not be either accumulated or blown-out by the winds in this disk. In addition, \citet{Riols18} claimed that such accumulation is absent in their two-dimensional simulations. Therefore, the MRI-driven disk winds may be insufficient to account for the inferred dust lifting.

Photoevaporation is an alternative mechanism to drive disk winds \citep{Owen11, Miotello12, Franz20, Hutchison21, Booth21}. \citet{Booth21} pointed out that delivery of grains to the disk ionization surfaces likely sets the maximum dust size entrained by the winds. This results in a factor of 10 reductions in size compared to the previous estimates \citep[e.g.,][]{Takeuchi05, Owen11, Franz20}. Nevertheless, micron-sized grains are possibly entrained by the winds even at a large radial distance, i.e., $100$ au, as long as a mass-loss rate is high enough \citep{Booth21}. Hence, it seems possible that the photoevaporation winds may play a role in lifting grains in the outer disk of HD 142527.

In Section \ref{sec:sizedist}, we suggested sub-micron-sized grains are underabundant at the disk surface. This seems to fit in the picture of dust entrainment by disk winds, where such tiny grains are efficiently removed. 

Disk winds in the outer disk of HD 142527 have been little constrained by observations so far. Kinematics of disk winds around T-Tauri stars have been studied with forbidden lines, such as [O I] and [Ne II] \citep[e.g.,][]{Hartigan95, Natta14, Rigliaco13, Simon16, Fang18, Banzatti19, Ballabio20, Pas20, Weber20, Whelan21}. However, for Herbig Ae/Be stars, disk winds are often not visible in atomic lines \citep{Acke05, Cauley14, Cauley15}. For HD 142527, \citet{Acke05} presented a [O I] spectrum, but they argued that its detection is confused by underlying photospheric absorption lines. Recently, \citet{Xu21} found a high-velocity wind component ($\sim250~\mathrm{km}~\mathrm{s}^{-1}$) in C II absorption lines, although this wind is likely being launched at the innermost region (<1 au).

Another possibility to probe disk winds is to observe molecular lines, such as CO emission lines \citep{Klaassen13, Teague19}. However, it also remains unclear whether the outer disk of HD 142527 exhibits disk winds, although \citet{Garg21} suggested a wind-like signature at the outermost regions ($\sim 600$ au). Thus, future observations probing gas kinematics of the outer disk would be helpful to elucidate the origin of dust lifting.

\section{Summary} \label{sec:summary}

We have performed radiative transfer simulations of the disks around HD 142527 to model the observed $3~\mu$m ice feature. The primary findings of this paper are as follows.

\begin{itemize}

\item The observed reddish scattered light spectra indicate the presence of grains with a few microns in size at the outer disk surface (Figure \ref{fig:spec}). Also, a narrow grain-size distribution seems favorable to explain the observations (Section \ref{sec:sizedist}). 

\item Dust grains at the disk surface have an ice/silicate mass ratio of $0.06-0.2$ regardless of a type of grain-size distribution. The inferred ice abundance is much lower than derived in \citet{Honda09}, and this is due to the isotropic scattering assumption in the previous study (Section \ref{sec:sizedist}). 

\item The inferred abundances are even lower than those estimated from far-IR emission features of water ice. This result points to the importance of ice disruption at the disk surface (Section \ref{sec:ice}).

\item The scattering feature profile depends on both ice abundance and grain-size distribution. It tends to peak at a wavelength around $2.9~\mu$m for ice-rich grains, while it peaks at a wavelength around $3.1~\mu$m for ice-poor grains (Section \ref{sec:prof}).

\item The ice feature strength depends on a scattering angle, particularly for micron-sized grains. As a consequence, the ice feature strength would not be uniform across the outer disk regions even if an ice abundance is the same throughout the disk (Figure \ref{fig:tauice}). Conversely, a different ice feature strength does not necessarily mean a different ice abundance.

\item A polarization fraction of disk-scattered light varies across the 3 $\mu$m feature as anticipated in previous studies. For small grains, a polarization fraction drops at $\lambda\sim3~\mu$m, whereas it is enhanced for micron-sized grains (Figure \ref{fig:polspec}). 

\item To explain the presence of micron-sized grains above the scattering surface, the disk should be subjected to either strong turbulence or disk winds. For the case of turbulent diffusion, $\alpha\sim2\times10^{-3}$ is at least necessary (Section \ref{sec:alpha}). Also, the current gas disk mass for HD 142527 based on CO emission observations is likely underestimated unless the actual turbulence is extremely strong ($\alpha\gtrsim0.3$) (Section \ref{sec:alphasum}).

\end{itemize}

We have also demonstrated that JWST/NIRCam is useful to study water ice for nearly face-on disks (Section \ref{sec:tauice}). Combined with NIRSpec for nearly edge-on disks, JWST will reveal water ice in disks at various inclination angles. Also, future observations probing gas kinematics at the outer disk surface of HD 142527 would be crucial to shed light on vertical grain dynamics.

\acknowledgments
We thank Christian Ginski, Carsten Dominik, and Hiroshi Kobayashi for their useful comments. We also thank Hidekazu Tanaka for making computational resources in his group available to us.  R.T. acknowledges JSPS overseas research fellowship. This work was supported by KAKENHI Grant Numbers JP17H01103 and JP18H05441.

\software{\texttt{RADMC-3D v2.0} \citep{Dullemond12}, \texttt{geofractal} \citep{Tazaki21}, \texttt{numpy} \citep{harris20}, \texttt{matplotlib} \citep{hunter07}.}

\appendix

\section{Effect of ice form on the near-IR ice feature in HD 142527} \label{sec:amo}

\begin{figure*}[t]
\begin{center}
\includegraphics[width=0.49\linewidth]{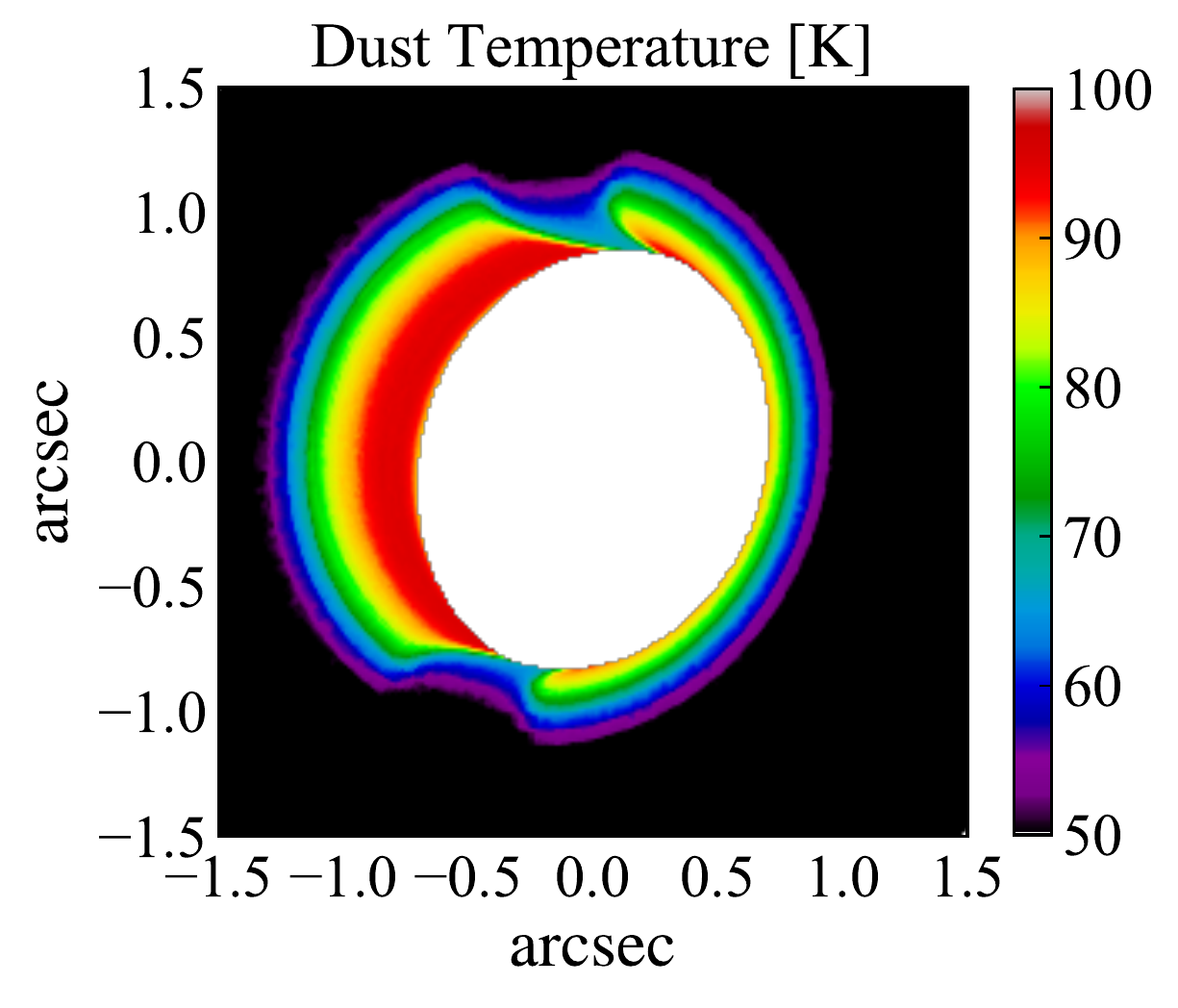}
\includegraphics[width=0.49\linewidth]{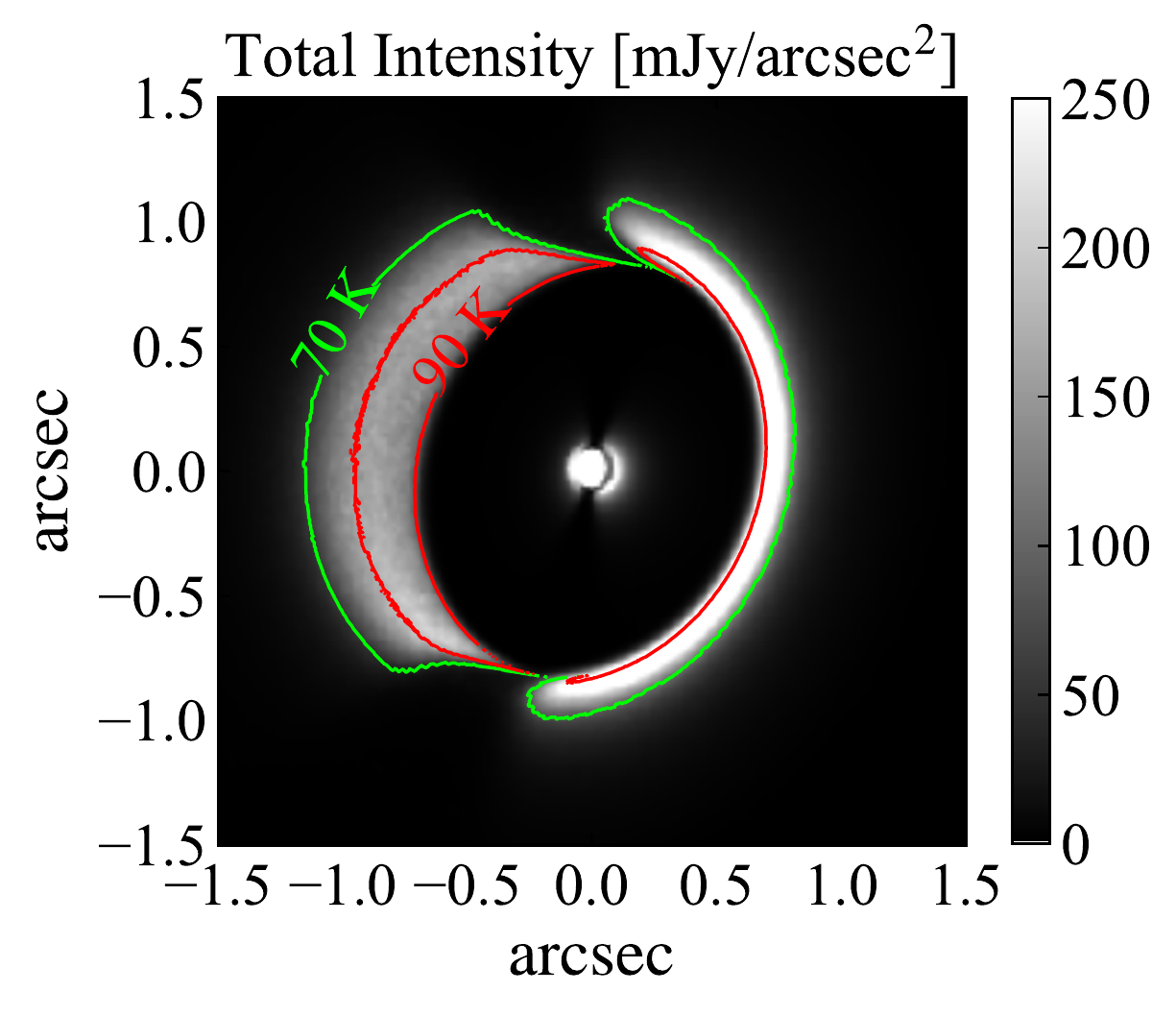}
\caption{Left and right panels show dust temperature at the scattering surface (only outer disk is shown) and total intensity, respectively. The assumed dust model is the log-normal distribution with $a_c=3~\mu$m and $\fice=0.1$ at $\lambda=2.2~\mu$m. The green and red contours in the right panel indicate the regions at which dust temperatures at the disk surface become 70 K and 90 K, respectively.}
\label{fig:td}
\end{center}
\end{figure*}

\begin{figure}[t]
\begin{center}
\includegraphics[width=\linewidth]{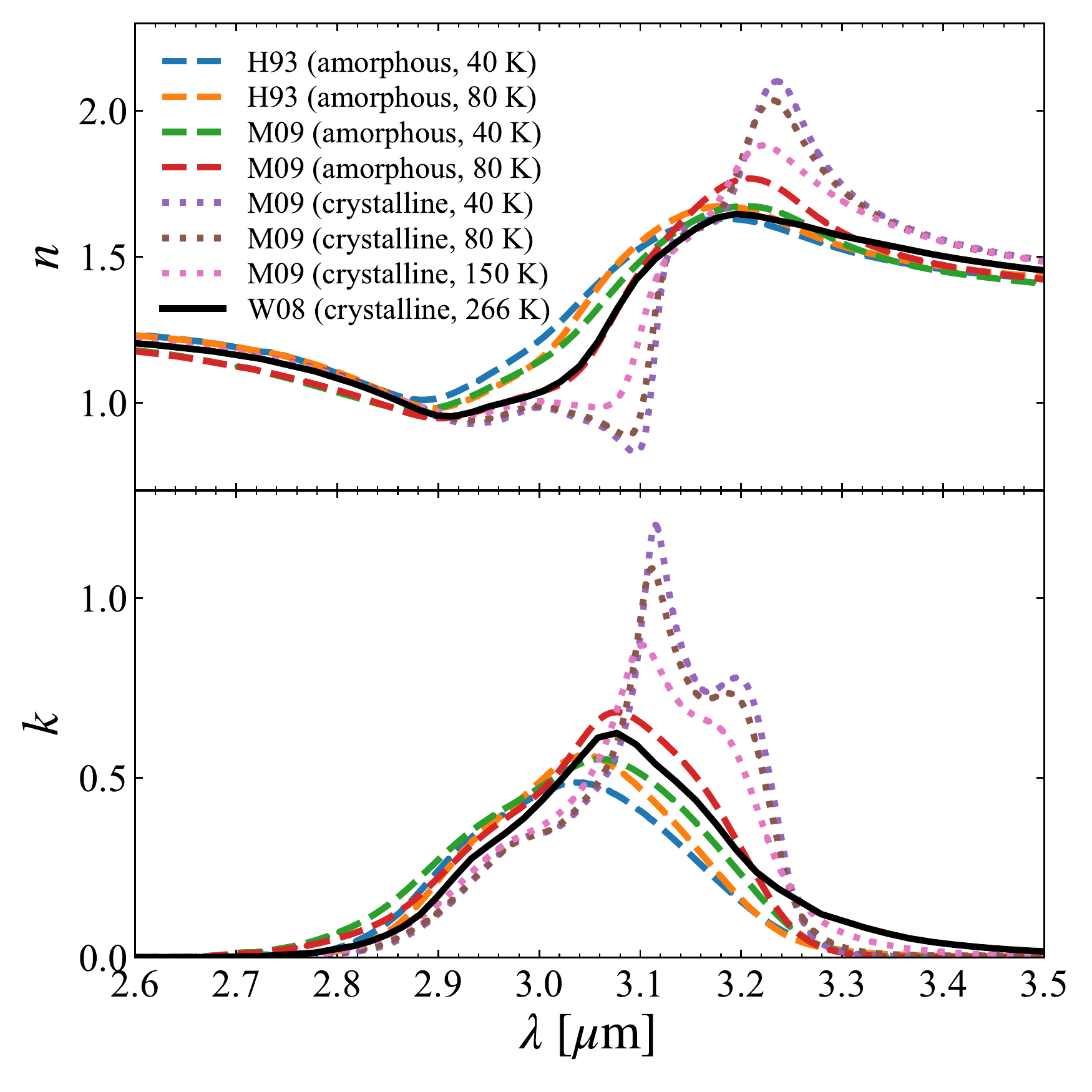}
\caption{Top and bottom panels show the real and imaginary part of the complex refractive index of water ice. The black solid line corresponds to crystalline ice in \citet{Warren08} (W08). The dashed and dotted lines correspond to amorphous and crystalline ice at various temperatures, respectively. The refractive indices are taken from \citet{Hudgins93} (H93) and \citet{Mastrapa09} (M09).}
\label{fig:opcont}
\end{center}
\end{figure}

This study adopted the refractive index proposed by \citet{Warren08} to model the observed $3~\mu$m feature of HD 142527. However, the refractive index at the 3-$\mu$m ice band depends on both temperature and form of water ice. Here we compare refractive indices of water ice under various conditions and discuss the impact on our results.

As the refractive index depends on temperature, we first assess ice temperature at the disk surface. Figure \ref{fig:td}(left) shows ice temperatures at the disk scattering surface at which the optical depth measured from the observer becomes unity. Also, Figure \ref{fig:td}(right) compares the temperature structure with a scattered light image. It turns out near-IR scattered light is mainly due to warm ice ($>70$ K). Although the inner wall of the outer disk is heated up above $90$ K, the near side at which \citet{Honda09} measured the scattering ice feature has a temperature of $\sim$80 K.

Figure \ref{fig:opcont} shows the refractive indices at 3-{\micron} ice band for amorphous and crystalline ice at various temperatures. The ice features are broad and single-peaked for amorphous ice, while they are sharp and triple-peaked for crystalline ice \citep[e.g.,][]{Bergren78, Hudgins93, Dartois01, Mastrapa09} (see also \citet{Whalley77} for the band assignment). In addition, amorphous and crystalline ice exhibit different temperature dependence. With increasing temperature, the amorphous ice features become narrower, stronger, and shifted to longer wavelengths, while the crystalline ice features become broader, weaker, and shifted shorter wavelengths \citep{Mastrapa09}. In particular, the triple peaks seen in crystalline ice are less clear for warmer ice.

Despite its crystalline nature, the $3~\mu$m feature of \citet{Warren08} does not show the triple peaks, presumably due to its high temperature \citep[266 K;][]{Schaaf73}. Instead, the ice band in \citet{Warren08} is close to that of amorphous ice with 80 K. Since this temperature is consistent with the near-side temperature (Figure \ref{fig:td}), the presence of amorphous ice at the disk surface would not change our results significantly. However, the crystalline ice band at 80 K is stronger than that of amorphous ice and \citet{Warren08}. This suggests that the presence of highly crystalline ice would make the ice abundance even smaller than the value estimated in this study.

\bibliography{cite}

\end{document}